\documentclass[12pt]{article}
\usepackage{setspace}
\usepackage{epic,eepic,epsfig,changebar,amsmath,sint}
\usepackage[american]{babel}

\newcommand{\fs}{\footnotesize}

\newdimen\arrayruleHwidth
\makeatletter
\def\Hline{\noalign{\ifnum0=`}\fi\hrule \@height \arrayruleHwidth
  \futurelet \@tempa\@xhline}
\makeatother

\def\rme{{\rm e}}
\def\rmO{{\rm O}}

\def\proof{\noindent{\sl Proof:}\kern0.6em}

\def\dual{\mathstrut^*\kern-0.1em}

\def\lvec#1{\setbox0=\hbox{$#1$}
    \setbox1=\hbox{$\scriptstyle\leftarrow$}
    #1\kern-\wd0\smash{
    \raise\ht0\hbox{$\raise1pt\hbox{$\scriptstyle\leftarrow$}$}}
    \kern-\wd1\kern\wd0}
\def\rvec#1{\setbox0=\hbox{$#1$}
    \setbox1=\hbox{$\scriptstyle\rightarrow$}
    #1\kern-\wd0\smash{
    \raise\ht0\hbox{$\raise1pt\hbox{$\scriptstyle\rightarrow$}$}}
    \kern-\wd1\kern\wd0}

\def\nabstar#1{\nabla\kern-0.5pt\smash{\raise 4.5pt\hbox{$\ast$}}
               \kern-4.5pt_{#1}}

\def\drvstar#1{\partial\kern-0.5pt\smash{\raise 4.5pt\hbox{$\ast$}}
               \kern-5.0pt_{#1}}

\def\MeV{{\rm MeV}}
\def\GeV{{\rm GeV}}

\def\fm{{\rm fm}}

\def\fpi{F_{\pi}}

\def\Nf{N_{\rm f}}
\def\psibar{\overline{\psi}}

\def\rhoprime{\rho\kern1pt'}
\def\rhobar{\bar{\rho}}
\def\rhobarprime{\rhobar\kern1pt'}
\def\rhobartilde{\kern2pt\tilde{\kern-2pt\rhobar}}
\def\rhobartildeprime{\kern2pt\tilde{\kern-2pt\rhobar}\kern1pt'}

\def\zetabar{\bar{\zeta}}
\def\zetaprime{\zeta\kern1pt'}
\def\zetabarprime{\zetabar\kern1pt'}
\def\zetar{\zeta_{\raise-1pt\hbox{\sixrm R}}}
\def\zetabarr{\zetabar_{\raise-1pt\hbox{\sixrm R}}}

\def\phiimpr{\phi_{\kern0.5pt\hbox{\sixrm I}}}

\def\diracstar#1#2{
    \setbox0=\hbox{$\gamma$}\setbox1=\hbox{$\gamma_{#1}$}
    \gamma_{#1}\kern-\wd1\kern\wd0
    \smash{\raise4.5pt\hbox{$\scriptstyle#2$}}}

\def\ca{c_{\rm A}}

\def\csw{c_{\rm sw}}

\def\ct{c_{\rm t}}

\def\ctildet{\tilde{c}_{\rm t}}

\def\fa{f_{\rm A}}

\def\fp{f_{\rm P}}

\def\f1{f_1}

\def\tr{\,\hbox{tr}\,}

\def\nf{N_{\rm f}}

\def\opprime#1{\setbox0=\hbox{${\cal O}$}\setbox1=\hbox{${\cal O}_{\rm #1}$}
    {\cal O}_{\rm #1}\kern-\wd1\kern\wd0
    \smash{\raise4.5pt\hbox{\kern1pt$\scriptstyle\prime$}}\kern1pt}

\def\ophatprime#1{\setbox0=\hbox{$\widehat{\cal O}$}
    \setbox1=\hbox{$\widehat{\cal O}_{\rm #1}$}
    \widehat{\cal O}_{\rm #1}\kern-\wd1\kern\wd0
    \smash{\raise4.5pt\hbox{\kern1pt$\scriptstyle\prime$}}\kern1pt}

\def\bopprime#1{\setbox0=\hbox{${\cal O}$}\setbox1=\hbox{${\cal O}_{\rm #1}$}
    {\cal L}_{\rm #1}\kern-\wd1\kern\wd0
    \smash{\raise4.5pt\hbox{\kern1pt$\scriptstyle\prime$}}\kern1pt}

\def\blagprime#1{\setbox0=\hbox{${\cal B}$}\setbox1=\hbox{${\cal B}_{#1}$}
    {\cal B}_{#1}\kern-\wd1\kern\wd0
    \smash{\raise5.2pt\hbox{\kern1pt$\scriptstyle\prime$}}\kern1pt}

\def\gbar{\bar{g}}
\def\mbar{\kern1pt\overline{\kern-1pt m\kern-1pt}\kern1pt}

\def\mq{m_{\rm q}}

\def\mr{m_{{\mbox{\scriptsize{\rm R}}}}}

\def\vbar{\bar{v}}

\def\zm{Z_{\rm m}}

\def\msbar{{\rm \overline{MS\kern-0.05em}\kern0.05em}}
\def\MSbar{{\rm \overline{MS\kern-0.05em}\kern0.05em}}

\begin{document}
\begin{titlepage}
\begin{flushright}
   DESY 04-217\\
   HU-EP-04-64\\
   MS-TP-04-31\\
   Bicocca-FT-04-16\\
   SFB/CPP-04-59\\
\today
\end{flushright}
\vskip 0.75 cm
\begin{center}
  {\Large\bf  Computation of the strong coupling in QCD with two
  dynamical flavours}
\end{center}
\vskip 0.5 cm
\begin{figure}[h]
\begin{center}
\epsfig{figure=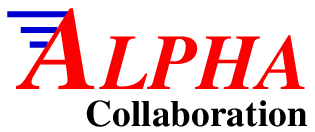} 
\end{center}
\end{figure}
\begin{center}
{\large     
M.~Della Morte\hskip .2ex$^{\scriptscriptstyle a}$,
R.~Frezzotti\hskip .2ex$^{\scriptscriptstyle b}$,
J.~Heitger\hskip .2ex$^{\scriptscriptstyle c}$,
J.~Rolf\hskip .2ex$^{\scriptscriptstyle a}$,
R.~Sommer\hskip .2ex$^{\scriptscriptstyle d}$
and
U.~Wolff\hskip .2ex$^{\scriptscriptstyle a}$}
\vskip 2.3ex
\vskip 1.0ex
$^{\scriptstyle a}$
Institut f\"ur Physik, Humboldt Universit\"at,
Berlin, Germany
\vskip 1.0ex
$^{\scriptstyle b}$
INFN--Milano and Universit\`a di Milano ``Bicocca'', Milano, Italy
\vskip 1.0ex
$^{\scriptstyle c}$
Institut f\"ur Theoretische Physik, Universit\"at M\"unster,
M\"unster, Germany
\vskip 1.0ex
$^{\scriptstyle d}$
DESY, Zeuthen, Germany

\vskip 1.0cm
{\bf Abstract}
\vskip 0.5ex
\end{center}
We present a non-perturbative computation of the running of the coupling
$\alpha_{\rm s}$ in QCD with two flavours of dynamical fermions in the
Schr\"odinger functional scheme. We improve our previous results by a 
reliable continuum extrapolation.
The $\Lambda$-parameter characterizing
the high-energy running is related to the value of the coupling
at low energy in the continuum limit. An estimate of 
$\Lambda r_0$ is given using large-volume data with
lattice spacings $a$ from $0.07\,\rm fm$ to $ 0.1\, \rm fm$. 
It translates into 
$\Lambda_\msbar^{(2)}=245(16)(16)\,\MeV$ [assuming $r_0 = 0.5\,\fm$]. 
The last step still has to be improved to reduce the uncertainty.
\vspace{.5cm}\\

 
\vfill \eject
\end{titlepage}

\section{Introduction}

The QCD sector of the Standard Model of elementary particles constitutes
a renormalizable quantum field theory. After determining one mass parameter
for each quark species and the strong coupling constant 
(at some reference energy) which fixes the interaction strength, 
QCD is believed -- and so-far found -- to predict all phenomena 
where only strong interactions are relevant. Due to asymptotic freedom,
this programme could be successfully implemented for processes with
energies $E \gg 1\,\GeV$ by using perturbation theory to evaluate QCD.
A recent report on such determinations of the coupling implied by a multitude
of experiments is found in \cite{Bethke:2004uy}.

At smaller energy the lattice formulation together with the tool of
numerical simulation is used to systematically extract predictions of the 
theory. 
Here the free parameters are typically determined by a sufficient number of 
particle masses or matrix elements associated with low energies, and then for 
instance the remaining hadron
spectrum becomes a prediction. See \cite{lat04:Ishikawa} for a recent 
reference on such large scale simulations. 

While this almost looks like two theories having disjoint domains of
applicability, we really view QCD as one theory for all scales. Then
the adjustable parameters mentioned above are not independent but half of
them should in fact be redundant or, in other words, predictable.
The ALPHA Collaboration is pursuing the long-term programme of computing
the perturbative parameters in the theory matched with Nature 
at hadronic energies. Here a non-perturbative multi-scale problem 
has to be mastered: hadronic and perturbative energies have to be covered
and kept remote from inevitable infrared and ultraviolet cutoffs.
To nevertheless gain good control over systematic errors in such a
calculation, a special strategy had to be developed; it will be reviewed 
in the next section.

As done for many other applications of lattice QCD, also our collaboration
has recently set out to overcome the quenched approximation and include
vacuum fluctuations due to
the two most important light quark flavours. In this article we
publish detailed results on the energy dependence of a non-perturbative
coupling from hadronic to perturbative energies and thereby connect the two 
regimes of QCD. This connection has been broken into two parts. It is first 
established with reference to a scale $L_{\rm max}$ in the hadronic sector 
that is not yet directly
physical but rather technically convenient within our non-perturbative
renormalization scheme. This part is, however, a universal continuum result
and represents what is mainly reported here. In a second step, which does not 
anymore involve large physical scale ratios, one number relating our 
intermediate result to physics, like for instance $L_{\rm max} F_{\pi}$, 
has to be computed. Here we can cite in this publication only a first 
estimate and not yet a systematic continuum extrapolation, which is left as 
a future task.

Earlier milestones of our programme consisted of the formulation of the 
finite-size scaling technique \cite{alpha:sigma}, 
its adaptation to QCD \cite{SF:LNWW,SF:stefan2},
check of universality \cite{alpha:SU2impr} and the complete
numerical execution of all steps in the quenched approximation 
\cite{mbar:pap1,mbar:pap3,lat01:jochen}.
A shorter account of the present study with a subset of the data
was published in~\cite{alpha:letter}.

We would like to mention that beside our finite-volume technique
there are also efforts to compute the coupling and quark masses 
in a more direct ``one-lattice'' approach. Some recent references
with unquenched results include
\cite{alpha:sesam,Gockeler:2004ad,alpha:davies03}. 
Since it is not possible to accommodate
all scales involved on one lattice in a satisfactory fashion, in these
works a perturbative connection is established between the bare coupling
associated with the lattice cutoff scale $a$ and a renormalized coupling
at high energy in a continuum scheme such as $\overline{\rm MS}$.
In our view, this step represents the main weakness of the 
method.\footnote{As for ref.~\cite{alpha:davies03}, an additional 
        relevant question concerns the correctness of the formulation of 
        fermions on the lattice. 
        We refer to \cite{Bunk:2004br} and references therein for an
        introduction to the problem.}
Series in the non-universal bare coupling are usually worse behaved
than renormalized perturbation theory. There are techniques to
improve this (tadpole improvement, boosted perturbation theory) which,
in cases where a non-perturbative check is possible, sometimes help
more and sometimes help less. Thus it is not easy to verify that one uses
this perturbation theory in a regime where it is accurate and to 
quote realistic errors on this step. 
On the other hand, this approach is much easier to carry through
and typically yields approximate results before an application
of our strategy is available.

\section{Strategy}

A non-perturbative renormalization of QCD addresses the question how
the high-energy regime, where perturbation theory has been successfully
applied in many cases, is related to the observed hadrons and their
interactions at small energies. This relation involves large scale
separations and thus is difficult to study by numerical
simulation. Naively, it would require simulations with a cutoff $a^{-1}$
much larger than the largest energy scale, combined with a large
system size $L$, much larger than the Compton wavelength of the
pion. In summary this would mean
\begin{equation}
  a^{-1} \gg  \mu_\text{perturbative} \gg \mu_\text{hadronic} \gg L^{-1}\,,
\end{equation}
to avoid discretization and finite-size errors. This clearly 
corresponds to -- with the current computing
power -- inaccessibly large lattices. 

Our method to overcome these problems has been developed
in~\cite{alpha:sigma,SF:LNWW,alpha:SU2,alpha:SU3,Sint:1994en,SF:stefan2,alpha:letter}; 
pedagogical introductions can be found 
in~\cite{reviews:schlad,reviews:leshouches2}. 
The key concept is an intermediate finite-volume 
renormalization scheme, in which the scale evolution of the coupling
(and the quark masses) can be computed recursively from low to very high
energies. At sufficiently high energies, the scale evolution is verified
to match with perturbation theory and there
the $\Lambda$-parameter is determined.

\subsection{Renormalization}

Any physical quantity $P$ should be independent of the renormalization
scale $\mu$. This is expressed by the Callan-Symanzik
equation~\cite{Callan:1970yg,Symanzik:1970rt} 
\begin{equation}
\label{eq:CS}
  \left\lbrace \mu\frac{\partial}{\partial\mu} +
  \beta(\gbar)\frac{\partial}{\partial\gbar} +
  \tau(\gbar)\mbar\frac{\partial}{\partial\mbar} \right\rbrace P = 0\,.
\end{equation}
Here, the $\beta$-function is given by 
\begin{equation}
\label{eq:ren}
  \beta(\gbar) = \mu\frac{\partial\gbar(\mu)}{\partial\mu}\,.
\end{equation}
Hence, once the coupling has been defined non-perturbatively for all
scales (see section~\ref{sec:SF}), also the $\beta$-function
is defined beyond perturbation theory.\footnote{For the $\tau$-function
  similar statements and expressions are 
  valid, once the running quark
  mass $\mbar$ has been defined non-perturbatively. 
} 
For weak couplings or
high energies only, the $\beta$-function can be asymptotically expanded as
\begin{equation}
\label{eq:expand}
  \beta(\gbar) = -\gbar^3 \left( b_0 + b_1 \gbar^2 + b_2 \gbar^4 +
  \ldots \right)\,.
\end{equation}
In the following we only consider mass independent renormalization
schemes~\cite{schemes:weinberg}, in which the renormalization conditions are 
imposed at zero quark mass. Particular examples are the Schr\"odinger
functional scheme described below, and the $\MSbar$ scheme. If in two
such schemes the couplings can\footnote{
See \cite{Frishman} for an example that this restriction can be non-trivial
for non-perturbative couplings.
}
be mutually expanded as Taylor series
of each other (once they are small enough),
\begin{equation}
\label{eq:twoschemes}
  \gbar'^{\,2} = \gbar^2 \left( 1+ C_g^{(1)} \gbar^2 + \ldots \right)\,,
\end{equation}
then the 1- and 2-loop coefficients $b_0$ and $b_1$ in (\ref{eq:expand}) 
are universal,
\begin{eqnarray}
  \label{eq:betauniversal}
  b_0 &=& \frac{1}{(4\pi)^2} \left( 11-\frac{2}{3}\Nf \right)\,, \\
  b_1 &=& \frac{1}{(4\pi)^4} \left( 102-\frac{38}{3}\Nf \right)\,,
\end{eqnarray}
while the higher-order coefficients depend on the choice of the
coupling. 
Starting from the 3-loop coefficient in the $\msbar$ 
scheme \cite{Tarasov:1980au}
and its conversion to the minimal subtraction scheme of lattice
regularization \cite{pert:LW95a,pert:LW95,threeloop:alles,Bode:2001uz}, 
the 3-loop  coefficient in 
the Schr\"odinger functional scheme (with all the parameters as
specified below) 
\begin{equation}
  \label{eq:betasf3loop}
  b_2 = \left(
        0.483(7) - 0.275(5)\Nf + 0.0361(5)\Nf^2 - 0.00175(1)\Nf^3
        \right)/(4\pi)^3
\end{equation}
could be obtained in~\cite{pert:2loop_fin}.

Now, a special exact solution of the Callan-Symanzik equation~(\ref{eq:CS})
is the renormalization group invariant $\Lambda$-parameter,
\begin{equation}
\label{eq:lambda}
  \Lambda = \mu\left(b_0\gbar^2(\mu)\right)^{-b_1/(2b_0^2)}
  \rme^{\,-1/(2b_0\gbar^2(\mu))} 
  \exp\left\lbrace -\int_0^{\gbar(\mu)} \! dx \left[ \frac{1}{\beta(x)} +
  \frac{1}{b_0x^3} - \frac{b_1}{b_0^2x} \right]\right\rbrace\,.
\end{equation}
$\Lambda$ is scale independent but renormalization scheme dependent. 
The transition to other mass independent schemes is accomplished exactly by 
a 1-loop calculation. If in another 
scheme the coupling is given through~(\ref{eq:twoschemes}) with
both $\gbar'$ and $\gbar$ taken at the same $\mu$,
the $\Lambda$-parameters are related by
\begin{equation}
  \Lambda' = \Lambda\,\rme^{\,C_g^{(1)}/(2b_0)}\,.
\end{equation}
In particular, the transition from the Schr\"odinger functional scheme
with two dynamical flavours
to the $\MSbar$ scheme of dimensional regularization  is
provided through~\cite{pert:1loop}
\begin{equation}
  \label{eq:lambdamsbar}
  \Lambda_\MSbar^{(2)} = 2.382035(3) \Lambda^{(2)}\,.
\end{equation}

\subsection{Schr\"odinger functional}
\label{sec:SF}

Since we want to connect the perturbative regime of QCD with the 
non-per\-tur\-bative hadronic regime, we have to employ a non-perturbative
definition of the coupling. Furthermore, the definition of the
coupling should be practical. This means that one has to be able to
evaluate it on the lattice with a small error, that cutoff effects are
reasonably small and that 
its perturbative expansion to 2-loop order is computable with a
reasonable effort. In principle there is a large freedom for the
choice of the coupling, however, it turns out that the conditions
above are hard to fulfill simultaneously.

To this end we consider the Schr\"odinger functional, which is the
propagation amplitude for going from some field configuration at the time
$x_0=0$ to another field configuration at the time $x_0=T$. 
Here the space-time is a hyper-cubic Euclidean lattice with discretization 
length $a$ and volume $T\times L^3$. We choose $T=L$ so that $L$ is the only
remaining external scale in the continuum limit $a\rightarrow 0$ of the 
massless theory.

The SU(3) gauge fields $U$ are defined on the
links of the lattice, while the fermion fields are defined on the
lattice sites. The partition function of this system is given by
\begin{equation}
  Z=e^{-\Gamma} = \int_{T\times L^3}\! {\rm D}[U,\psi,\psibar]\
  e^{-S[U,\psi,\psibar]}\,. 
\end{equation}
Here, the action is the sum 
$S[U,\psi,\psibar] = S_{\rm g}[U] + S_{\rm f}[U,\psi,\psibar]$ 
of the $\rmO(a)$ improved plaquette action
\begin{equation}
  S_{\rm g}[U] = \frac{1}{g_0^2} \sum_p w(p) \tr (1-U(p))
\end{equation}
and the fermionic action
\begin{equation}
  S_{\rm  f}[U,\psi,\psibar] = \sum_x \psibar(x) (D+m_0)\psi(x)
\end{equation}
for two degenerate flavours implicit in $\psi$. 
For the special boundary conditions considered below,
the weight factor $w(p)$ is the boundary improvement term
$\ct$~\cite{SF:LNWW} for time-like plaquettes at the boundary and one in
all other cases. The value of $\ct$ has become available to 2-loop order in
perturbation theory~\cite{pert:2loop_fin} in the course of this
work. Thus some of our data sets use the 1-loop value of
$\ct$, hence our simulations adopt two different actions. 
Because of universality we expect them to yield the same continuum limits 
for our observables.

The $\rmO(a)$ improved Wilson Dirac operator $D$ includes the
Sheikholeslami-Wohlert term \cite{impr:SW} multiplied with the improvement
coefficient $\csw$ that has been determined with non-perturbative
precision in~\cite{impr:csw_nf2}, and a boundary improvement term that is
multiplied by the coefficient $\ctildet$, which is known to 1-loop order. 
For details and notation we refer to~\cite{impr:pap1,impr:pap2}. 

The boundary conditions in the space directions are periodic for the
gauge fields and periodic up to a global phase $\theta$
for the fermion fields. The value of $\theta$ was optimized at 1-loop
order of perturbation theory~\cite{pert:1loop}. It turns out
that a value close to $\theta=\pi/5$ leads to a significantly smaller
condition number of the fermion matrix than other values of $\theta$
and thus to a smaller 
computational cost. Benchmarks in the relevant parameter range for our
project have shown this too, and therefore we adopt the choice
$\theta=\pi/5$ in this work.

In the time direction, Dirichlet boundary conditions are imposed at
$x_0=0$ and $x_0=T$. The quark fields at the boundaries are given by
the Grassmann valued fields $\rho, \bar{\rho}$ at $x_0=0$ and $\rho',
\bar{\rho}'$ at $x_0=T$, respectively. They are used as sources that
are set to zero after differentiation. The gauge fields at the
boundaries are chosen such that a constant colour-electric background
field, which is the unique (up to gauge transformations) configuration
of least action, is generated in our space-time~\cite{SF:LNWW}. 
This is achieved by the diagonal colour matrices specified in
ref.~\cite{alpha:SU3}, parametrized by two dimensionless real 
parameters $\eta$ and $\nu$.

A renormalized coupling $\gbar^2$ may then
be defined by differentiating the effective action $\Gamma$ at the
boundary point ``A'' of~\cite{alpha:SU3} that corresponds to the choice
$\eta=\nu=0$,
\begin{equation}
  \label{eq_Gammap}
  \frac{\partial \Gamma}{\partial\eta}\bigg\vert_{\eta=\nu=0} =
  \frac{k}{\gbar^2}\,. 
\end{equation}
The normalization
$k$ is chosen such that the tree-level value of $\gbar^2$ equals
$g_0^2$ for all values of the lattice spacing. The boundary point ``A''
and especially the value $\nu=0$ are used, since the statistical error
of the coupling turns out to be small for this choice. For general
values of $\nu$ we find another renormalized quantity $\vbar$,
\begin{equation}
  \label{eq:vbar}
  \frac{\partial \Gamma}{\partial\eta}\bigg\vert_{\eta=0} =
  k\left\lbrace \frac{1}{\gbar^2}-\nu\vbar\right\rbrace\,,
\end{equation}
that we have investigated as well to study the effects of dynamical
fermions.

The renormalized coupling depends on the system size, the
lattice spacing $a$ and on the
quark mass. The bare quark mass $m_0$ is additively renormalized on
the lattice because chiral symmetry is broken for Wilson 
fermions~\cite{Wilson}.
Thus we define the bare mass by the PCAC relation that
relates the axial current $A^a_\mu(x) =
\psibar(x)\frac{\tau^a}{2}\gamma_\mu\gamma_5\psi(x)$ to the pseudoscalar 
density
$P^a(x) = \psibar(x)\frac{\tau^a}{2}\gamma_5\psi(x)$. 
Using the matrix elements $\fa$ and $\fp$ of $A^a_\mu$ and $P^a$, 
respectively (cf.~\cite{impr:pap1}), the $\rmO(a)$ improved PCAC mass is 
defined as
\begin{equation}
  m(x_0) = \frac{\frac{1}{2}(\partial_0+\partial_0^{\star}) \fa(x_0) +
  \ca a \partial_0^{\star}\partial_0\fp(x_0)}{2\fp(x_0)}\,.
\end{equation}
We have used 1-loop perturbation theory for $\ca$~\cite{impr:pap2}.

To fix all the details of our scheme, we define the bare current mass
through 
\begin{equation}
  m = \left\lbrace
    \begin{array}{ll}
      m(T/2)\,, & \mbox{if $T/a$ is even,}\\
      \left[m((T-a)/2) + m((T+a)/2)\right]/2\,, & \mbox{if $T/a$ is odd.}
    \end{array}\right.
\end{equation}
This mass is tuned to zero, 
\begin{equation}
  \label{eq:massenull}
  m(g_0,m_0,L/a) = 0\,,
\end{equation}
so that we have a massless renormalization scheme, in which the only 
remaining external scale in the continuum limit is the system size $L$.

\subsection{Computational Strategy}

In the last section we have defined the Schr\"odinger functional
coupling $\gbar^2$. 
This finite-volume coupling runs with
$\mu=1/L$ and -- assuming monotonicity -- there is a one-to-one 
relation between the value of the
coupling and the system size $L$ or energy scale $\mu$. In an
abuse of notation we will from now on write $\gbar^2(L)$ instead of 
$\gbar^2(1/L)$.

Our goal is to calculate the scale evolution of the strong coupling and
the $\text{$\Lambda$-parameter}$ of QCD in terms of a low-energy scale.
We start the computation by choosing a value $u$ for the renormalized
coupling (which implicitly determines $L$) and by choosing a
lattice resolution $L/a$. The theory can 
then be renormalized by tuning the bare parameters $\beta = 6/g_0^2$
and $\kappa = 1/(8+2am_0)$
such that 
\begin{equation}
\label{eq:tuning}
  \gbar^2(L) = u \quad\text{and}\quad m = 0\,.
\end{equation}
Now we simulate a lattice
with twice the linear size at the same bare parameters, that means at
the same value of the lattice spacing, and thus with the physical
extent $2L$, corresponding to a new renormalized coupling
$u'=\gbar^2(2L)$.\footnote{The mass $m$ on this larger lattice is different 
from zero by a lattice artifact which is expected to vanish in the
continuum limit proportionally to $a^2$.
This has been verified in ref.~\cite{alpha:letter}.}
This determines the scale evolution of the renormalized
coupling. It can be expressed through the lattice step scaling function
\begin{equation}
  \left.\Sigma(u,a/L) = \gbar^2(2L)\right\vert_{\gbar^2(L)=u\,,\,m=0}\,,
\end{equation}
which is the key observable we compute.
Finally, we obtain the step scaling function
\begin{equation}
  \sigma(u) = \lim_{a/L\rightarrow 0}\Sigma(u,a/L)
\end{equation}
in the continuum limit by repeating these three steps 
with finer and finer lattice resolutions.
This algorithm is iterated for a sequence of values for $u$ to get the
functional form of $\sigma(u)$.

For small values of $u$ the step scaling function $\sigma(u)$ 
can be expanded in renormalized perturbation theory,
\begin{equation}
\label{eq:sigmaPT}
  \sigma(u) = u + s_0 u^2 + s_1 u^3 + \ldots\,,
\end{equation}
with the coefficients
\begin{eqnarray}
  \label{eq:scoeff}
  s_0 &=& 2\ b_0 \ln 2\,,\\
  s_1 &=& (2\ b_0 \ln 2)^2 +  2\ b_1 \ln 2\,,\\
  s_2 &=& (2\ b_0 \ln 2)^3 + 10\ b_0\ b_1 (\ln 2)^2 + 2\ b_2 \ln 2\,.
\end{eqnarray}
The step scaling function $\sigma(u)$ can be interpreted as an
integrated discrete $\beta$-function. Indeed, by using
equation~({\ref{eq:ren}}) we get
\begin{equation}
  \label{eq:betarec}
  \beta\left(\sqrt{\sigma(u)}\right) = \beta\left(\sqrt{u}\right)\,
  \sqrt{\frac{u}{\sigma(u)}}\,\sigma'(u)
\end{equation}
for the $\beta$-function, which allows to calculate it recursively once the 
step scaling function $\sigma(u)$ is known. 

To arrive at our main result, that is the $\Lambda$-parameter in terms of
a low-energy scale, we solve the equation
\begin{equation}
\label{eq:rec}
  \sigma\left(\gbar^2(L/2)\right) = \gbar^2(L)
\end{equation}
recursively for $\gbar^2(L/2)$. We start this recursion at a maximal
value $u_{\rm max}=\gbar^2(L_{\rm max})$ of the coupling. The value of
$u_{\rm max}$ is chosen such that the associated scale $L_{\rm max}$
is a scale in the hadronic regime of QCD. Following the
recursion~(\ref{eq:rec}) to larger and larger energies, we obtain the
values for
\begin{equation}
  u_i = \gbar^2(2^{-i}L_{\rm max})\,,\quad i=0,\ldots,n\,,
  \quad u_0=u_{\rm max}\,.
\end{equation}
We perform $n=7$ or $n=8$ steps of this recursion and can in this way
cover a scale separation of a factor 100 to 250. 
Eventually, for sufficiently large energies, perturbation theory can safely 
be applied. 
Then we use~(\ref{eq:lambda}) with $\mu=2^n/L_{\rm max}$ and with the
$\beta$-function truncated at 3-loop order,
(\ref{eq:betauniversal})--(\ref{eq:betasf3loop}). The final result for
$\Lambda L_{\rm max}$ in the Schr\"odinger functional scheme can be
converted to the $\MSbar$ scheme with~(\ref{eq:lambdamsbar}). We also check
the admissibility of employing perturbation theory by studying the variation 
of our final result with respect to the number of non-perturbative steps
$n$ in the scale evolution of the strong coupling. 

\subsection{Discretization effects}

The influence of the underlying space-time lattice on the evolution of
the coupling can be estimated perturbatively~\cite{pert:2loop_fin}, by
generalizing 
Symanzik's discussion~\cite{Symanzik:1982,impr:Sym1,impr:Sym2} to the
present case. Close to the continuum 
limit we expect that the relative deviation 
\begin{equation}
  \delta(u,a/L) = \frac{\Sigma(u,a/L)-\sigma(u)}{\sigma(u)} = 
  \delta_1(a/L) u + \delta_2(a/L) u^2 + \ldots
\end{equation}
of the lattice step scaling
function from its continuum limit converges to zero with a rate
roughly proportional to $a/L$. 
More precisely, since the action is $\rmO(a)$ improved, we expect
\begin{eqnarray}
  \delta_1(a/L) &\sim& \left(d_{0,1} + d_{1,1}\ln\frac{a}{L}\right)
  \left(\frac{a}{L}\right)^2 + \ldots\,,\\
  \delta_2(a/L) &\sim& e_{0,2} \frac{a}{L}  + \left(d_{0,2} +
  d_{1,2}\ln\frac{a}{L}+
  d_{2,2}\left(\ln\frac{a}{L}\right)^2\right) 
  \left(\frac{a}{L}\right)^2 + \ldots
\end{eqnarray}
for the 1-loop value of $\ct$ and the same form with $e_{0,2}=0$ for
the 2-loop value of $\ct$.
Note that the tree-level discretization
effects vanish exactly, since we normalize the coupling such that its
perturbative expansion starts with $g_0^2$ for all values of the
lattice spacing. 

The coefficients $\delta_1$ and $\delta_2$ are collected in
table~\ref{tab:ssfdisc} for the resolutions needed in this work.
\begin{table}[htbp]
  \centering
  \begin{tabular}{clll}
    \hline\\[-1.5ex]
    $L/a$ & $\delta_1$ & $\delta_2^{\ct=\text{1-loop}}$ &
    $\delta_2^{\ct=\text{2-loop}}$\\[0.5ex] \hline\\[-1.5ex]
    4 & $-0.0103$ & 0.0063 & $-0.00007$\\
    5 & $-0.0065$ & 0.0049 & $-0.00019$\\
    6 & $-0.0042$ & 0.0038 & $-0.00041$\\
    8 & $-0.0021$ & 0.0029 & $-0.00030$\\[0.5ex] \hline
  \end{tabular}
  \caption{\fs Discretization error of the step scaling function.}
  \label{tab:ssfdisc}
\end{table}
An expanded version of this table can be found in~\cite{alpha:bermions}.
The entries in the last column are very small. For larger values of
$L/a$ than shown in the table, $\delta_2^{\ct=\text{2-loop}}$ decreases as 
expected. 
Since $\delta_2^{\ct=\text{1-loop}}$ is of the order $a/L$, it is no
surprise that it is much larger than $\delta_2^{\ct=\text{2-loop}}$.
In fact, it is of the same size as $\delta_1$, for which the linear term 
in $a/L$ is absent.

The largest coupling at which the step scaling function has been
computed with the 1-loop value of $\ct$ is $u=1.7319$. With the
2-loop value of $\ct$, this is $u=3.334$. Table~\ref{tab:ssfdisc}
suggests that the step scaling function is only mildly affected by
discretization effects. This will be demonstrated by our numerical
results in section~\ref{sec:results}.

We cancel the known perturbative cutoff effects for the respective
actions by using 
\begin{equation}
\label{eq:Sigma2}
  \Sigma^{(2)}(u,a/L) = \frac{\Sigma(u,a/L)}{1 + \delta_1(a/L) u +
  \delta_2(a/L) u^2}
\end{equation}
in the analysis of our Monte Carlo data. The perturbative estimate 
of the relative cutoff effects behaves as $(a/L)\times u^3$
close to the continuum limit.

\subsection{Matching to a hadronic scheme \label{sect:hadronic}}

As described so far, 
our computational strategy yields $\Lambda L_{\rm max}$,
with $L_{\rm max}$ defined by the value of the coupling itself. Since the
latter is not experimentally measurable, it remains to relate $L_{\rm max}$
to a hadronic scale. Here, the natural choice is 
the pion decay constant $F_\pi$, since chiral
perturbation theory provides an analytic understanding of the 
pion dynamics \cite{chir:Weinberg,chir:GaLe1}, which 
is expected to help to control the extrapolations to the physical 
quark mass \cite{chir:GaLe1} as well as to infinite 
volume \cite{lat04:gilberto}.

A computation of $L_{\rm max}F_\pi$ requires the knowledge
of $f(\tilde{g}_0^2)=aF_\pi$ (at a quark mass where $m_\pi/F_\pi$ takes
its experimental value) and $l(g_0^2)=L_{\rm max}/a$, where 
$\gbar^2(L_{\rm max})=u_{\rm max}$. 
We remind the reader that $\gbar^2$ is defined at vanishing quark mass. 
The difference between the improved bare coupling 
$\tilde{g}_0$ \cite{impr:pap1} and $g_0$ is proportional 
to the light quark mass and can safely be neglected for
physical values of the light quark mass. We therefore replace
$\tilde{g}_0 \approx g_0$ in the following. 
The value of $u_{\rm max}$
is restricted to be in the range covered by the computation
of the scale dependence of the coupling and, for lattice
spacings accessible in large-volume simulations,
$l(g_0^2)$ should be sufficiently large. With both functions,
$l$ and $f$, defined for the same discretization, one finally
wants to evaluate  
\begin{equation}
  \left.  L_{\rm max} \,F_\pi \right|_{\rm continuum} =
  \lim_{f \to 0} l(g_0^2) f(g_0^2) \,.
\end{equation}
Unfortunately, the  results available in the literature 
\cite{UKQCD:nf2b52b,JLQCD:nf2b52} 
for $f(g_0^2)$ with our action~\cite{impr:csw_nf2} suffer from an 
uncertainty in the renormalization of the axial current, 
which has not yet been performed non-perturbatively. 
Also the O$(a)$ improvement of the current is known only perturbatively
and it is not obvious that quark masses have been reached 
where chiral perturbation theory is applicable.

At present, we thus prefer to relate $L_{\rm max}$ to the frequently used 
hadronic radius $r_0$, which, according to phenomenological
considerations, has a value of around $0.5\,\fm$~\cite{pot:r0} and which 
has also been the reference scale in the zero-flavour theory, 
i.e.~the quenched approximation~\cite{mbar:pap1}. 
Note that in a calculation in this approximation agreement was found
(within its 3\% precision) between $F_{\rm K} r_0$ and the product of the 
experimental number for $F_{\rm K}$ times $0.5\,$fm \cite{mbar:pap3}. 
Below, we thus evaluate $L_{\rm max} /r_0 =  l(g_0^2) / \rho(g_0^2)$ where
$\rho(g_0^2) = r_0/a$ and translate to physical units via $r_0=0.5\,\fm$.


\section{Details of the numerical simulation and analysis}

In addition to the piece
present in the pure gauge theory \cite{alpha:SU3},
the central observable, $\partial \Gamma/\partial \eta$,
receives a contribution due to the quark determinant. 
We describe its numerical evaluation by a stochastic estimator
in appendix~\ref{sect:dsdeta}. Also detailed tables with
simulation results are deferred to an appendix (\ref{sect:res}). 

\subsection{Tuning}

With a number of tuning runs we determine the bare parameters $\beta$
and $\kappa$ such that eq.~(\ref{eq:tuning})
is valid.
Fulfilling the condition $m(L)=0$ precisely would require a fine
tuning of the hopping parameter $\kappa$. 
However, the uncertainty in
$\Sigma$ owing to a small mismatch of $m$ can be estimated
perturbatively. To this end we compute the derivative of $\Sigma$ with
respect to $z=mL$,
\begin{equation}
  \left.\frac{\partial}{\partial z} \gbar^2(2L)
  \right\vert_{\gbar^2(L)=u\,,\,m(L)=z/L} = \Phi(a/L) u^2 + \ldots \,.
\end{equation}
It turns out that $\Phi$ is a slowly varying function of
$a/L$. Thus, for our purpose it suffices to approximate it by its
universal part,
\begin{equation}
  \Phi(0) = -\frac{\Nf}{4\pi} \left. \frac{\partial}{\partial z}
  c_{1,1}(z)\right\vert_{z=0} = 0.00957 \Nf \,,
\end{equation}
where $c_{1,1}(z)$ has been taken from~\cite{pert:1loop}. The typical
precision in Monte Carlo simulations is $\Delta(\gbar^{-2}) =
\gbar^{-4}\Delta(\gbar^2) = 0.003$, as can be seen from the
tables~\ref{tab:data1} and~\ref{tab:data2} in appendix~\ref{sect:res}. 
This means that an additional error of $0.001\times u^2$ due to a slight 
mismatch of the mass $m(L)$ is tolerable. Then it suffices to require
\begin{equation}
  am < 0.1 \frac{a}{L} \frac{1}{\Nf}\,.
\end{equation}
The tables~\ref{tab:data1} and~\ref{tab:data2} in appendix~\ref{sect:res} 
show that we have reached this precision in our simulations.

Analogously, we stop the fine tuning of $\beta$ if $\gbar^2(L)=u$
well within the errors. Then we can correct for a small mismatch owing to 
$\gbar^2(L)=\tilde{u}\neq u$ by using
\begin{equation}
  \Sigma(u,a/L) = 
  \Sigma(\tilde{u},a/L) + \Sigma'(u,a/L)\times(u-\tilde{u})\,,
\end{equation}
with the perturbative estimate
\begin{equation}
  \Sigma'(u,a/L)\equiv\frac{\partial\Sigma(u,a/L)}{\partial u} \approx
  \frac{\partial\sigma(u)}{\partial u} \approx 1 + 2 s_0 u + 3 s_1
  u^2 + 4 s_2 u^3
\end{equation}
for the derivative of the step scaling function $\Sigma$. This
correction is always smaller than the statistical error of the step scaling
function.

Similarly, we convert the statistical error on $u$ into an additional error of
$\Sigma$,
\begin{eqnarray}
  \Delta\left(\Sigma(u)\right)  \approx 
  \Sigma'(u,a/L)\times\Delta(u) \approx 
  (1 + 2 s_0 u + \ldots) \times \Delta(u) \,.
\end{eqnarray}
This additional error is always much smaller than the error of $\gbar^2(2L)$,
to  which it is added in quadrature.

In some cases we have stopped the fine tuning of $\beta$ and $\kappa$ after
a number of runs and interpolated the results such that exactly the target
coupling and mass zero with the errors as shown in
the tables~\ref{tab:data1} and~\ref{tab:data2} were obtained.

The step scaling function $\Sigma(u,a/L)$ and its partner with the 
perturbative cutoff effects being divided out, $\Sigma^{(2)}(u,a/L)$
(cf.~(\ref{eq:Sigma2})), are listed in table~\ref{tab:Sigma}.
\begin{table}[htbp]
  \centering
  \begin{tabular}{llllllll} \hline\\[-1.5ex]
$u$ &$L/a$ & $\Sigma(u,a/L)$ &$\Sigma^{(2)}(u,a/L)$ & $u$ &$L/a$ & $\Sigma(u,a/L)$ &$\Sigma^{(2)}(u,a/L)$ \\[0.5ex] \hline\\[-1.5ex]
0.9793 & 4 & 1.0643(35) & 1.0686(35)    &         1.5031 & 4 & 1.7204(56) & 1.7477(57)\\      
  & 5 & 1.0720(40) & 1.0738(41)         &           & 5 & 1.737(11) & 1.755(11)\\             
  & 6 & 1.0802(46) & 1.0807(46)         &           & 6 & 1.730(13) & 1.743(13)\\             
  & 8 & 1.0736(59) & 1.0729(59)         &           & 8 & 1.723(16) & 1.730(16)\\[0.75ex]      
1.1814 & 4 & 1.3154(55) & 1.3199(56)    &         2.0142 & 4 & 2.481(18) & 2.535(18)\\        
  & 5 & 1.3296(61) & 1.3307(61)         &           & 5 & 2.438(20) & 2.473(20)\\             
  & 6 & 1.3253(70) & 1.3249(70)         &           & 6 & 2.507(27) & 2.533(28)\\             
  & 8 & 1.3342(71) & 1.3323(71)         &           & 8 & 2.475(35) & 2.489(35)\\[0.75ex]      
1.5031 & 4 & 1.7310(61) & 1.7332(61)    &         2.4792 & 4 & 3.251(28) & 3.338(29)\\        
  & 5 & 1.756(12) & 1.754(12)           &           & 5 & 3.336(52) & 3.394(53)\\             
  & 6 & 1.745(12) & 1.741(12)           &           & 6 & 3.156(57) & 3.198(58)\\             
  &   &           &                     &           & 8 & 3.326(52) & 3.351(53)\\[0.75ex]      
1.7319 & 4 & 2.0583(76) & 2.0562(76)    &         3.334 & 4 & 5.588(54) & 5.791(56)\\         
  & 5 & 2.083(21) & 2.076(21)           &           & 5 & 5.43(11) & 5.56(11)\\               
  & 6 & 2.058(20) & 2.049(20)           &           & 6 & 5.641(99) & 5.75(10)\\              
  &   &           &                     &           & 8 & 5.48(13) & 5.53(13)\\[0.5ex] \hline        
  \end{tabular}
  \caption{\fs Step scaling functions $\Sigma$ and $\Sigma^{(2)}$. The
    left hand side of the table contains the data with the 1-loop
    value of $\ct$, while the data with $\ct=\text{2-loop}$ are shown on the 
    right.}
  \label{tab:Sigma}
  \vspace{-0.5cm}
\end{table}

\subsection{Parameters}

Our choice of parameters is displayed in table~\ref{tab:data1} and
table~\ref{tab:data2} in appendix~\ref{sect:res}. 
The parameters shown are the results of a careful application of the 
tuning procedure explained in the last section. 
The tables reveal that the condition $\gbar^2(L)=u$ is
fulfilled to a good precision. The remaining deviations and the errors
are then propagated into an additional error for $\Sigma(u)$, as
described above.

\subsection{Simulation costs and proper sampling of the 
configuration space}
\label{sec:algo}

Most of our results have been produced with the hybrid Monte Carlo
algorithm (HMC)~\cite{Duane:1987de}, the polynomial hybrid Monte Carlo
(PHMC) in the version proposed in~\cite{phmc:pap1,phmc:pap2} and for
some of the $L/a=16$ runs the
hybrid Monte Carlo algorithm generalized to two pseudofermion
fields (HMC~2~pf)~\cite{algo:GHMC,Hasenbusch:2002ai}. 

We measure the cost of our simulations with the quantity
\begin{eqnarray}
M_{\rm cost} & = & {\rm(update \; time \; in \; seconds \; on \;
  machine \; M}) \\ \nonumber
&& \times \, {\rm{(error \; of}} \; 1/\bar{g}^2)^2 \times (4a/T)(4a/L)^3 \,.
\end{eqnarray}
In~\cite{alpha:bench} the cost of a subset of the simulations discussed here
(essentially up to $L/a=12$ and $\gbar^2\approx 2.5$) has been
analyzed. A typical value for HMC at $L/a=10$ and
$\gbar^2\approx2.5$ is $M_{\rm cost} \approx 3.5$, with respect to one board
of APEmille \cite{APEmille_lat01}. 
To give an idea about the increase of the cost for the $L/a=16$
simulations, we collect results from two different runs at the coupling 
$\bar{g}^2\approx 2.5$ in table~\ref{mcost_tab}.
We compute the autocorrelation time $\tau_{\rm int}$ in units of
trajectories in the way suggested in~\cite{Wolff:2003sm}. 
\begin{table}[htb]
  \begin{center}
    \begin{tabular}{ccccccc}
      \hline\\[-1.5ex]
      algorithm  &  step size & $\tau_{\rm int}$   & $t_{\rm
        update}/[\,s\,]$ & $\gbar^2$ & $N_{\rm traj} $ & $M_{\rm cost}$ \\[0.5ex]
      \hline\\[-1.5ex]
      HMC & $0.0625$ & $3.6(3)$ & $380$ & $2.46(5)$ & $4800$ & $6.9(6)$ \\
      HMC~2~pf & $0.111$  & $4.6(4)$ & $262$ & $2.55(5)$ & $5900$ & $5.6(5)$ \\[0.5ex]
      \hline
    \end{tabular}
    \caption{\fs Cost estimates for two $L/a=16$ runs; 
      $t_{\rm update}/[\,s\,]$ is the time in seconds needed for one
      trajectory of length 1. The reference machine is an APEmille board.
\label{mcost_tab}}
  \end{center}
\vspace{-0.5cm}
\end{table}

Related to the issue of estimating the autocorrelation time
is always the question whether the algorithm samples the 
entire relevant configuration space efficiently. If this is not the case, 
autocorrelation times may be largely underestimated  and even 
systematically wrong
results may be obtained. We now discuss two at least rough checks 
that our simulations
do not suffer from such problems. 

1. To investigate the contributions to the coupling from sectors of
configuration space with non-trivial topology, we have performed a set
of simulations at the coupling $\gbar^2\approx 2.5$ with $L/a=8,12$. 
The topological charge $Q(U)$ is determined through a cooling procedure.
Since sampling different topological sectors might
be algorithmically very difficult, we have employed the PHMC algorithm,
whose flexibility can be exploited to enhance the transition rate among
different sectors (at the price of increasing the fluctuations of the
reweighting factor), and checked the results to be
independent of the polynomial approximation used.

Starting from a hot, random configuration, several of the independent
replica were in a non-trivial topological sector. For these replica
the smallest eigenvalue of $D^2$ (even-odd preconditioned) turns out
to be one order of magnitude smaller than typical values in the
topologically trivial sectors. After $\rmO(100)$ trajectories for
$L/a=8$, respectively O(1000) trajectories for $L/a=12$, all the
replica have zero topological charge and transitions to sectors with
$Q(U)\neq 0$ have not been observed in additional $\rmO(10^4)$ trajectories.

From this we conclude that the PHMC algorithm can tunnel between different
topological sectors, but for large $L/a$ the transition rate is
very small. In addition, the weight of the non-trivial sectors is
too small for them to occur in a practical simulation at all (all
tunnellings went to $Q(U)=0$ and none in the reverse direction). Their
weight in the path integral is negligible. 
These statements have been checked for $L/a=8,12$, and
it appears safe to assume their validity also for larger
$L/a$. Therefore, we decided to
always start from a cold configuration, especially for the $L/a=16$
simulations, to avoid thermalization problems.

\begin{figure}[htb]
  \begin{center}
\vspace{-1.5cm}
    \epsfig{file=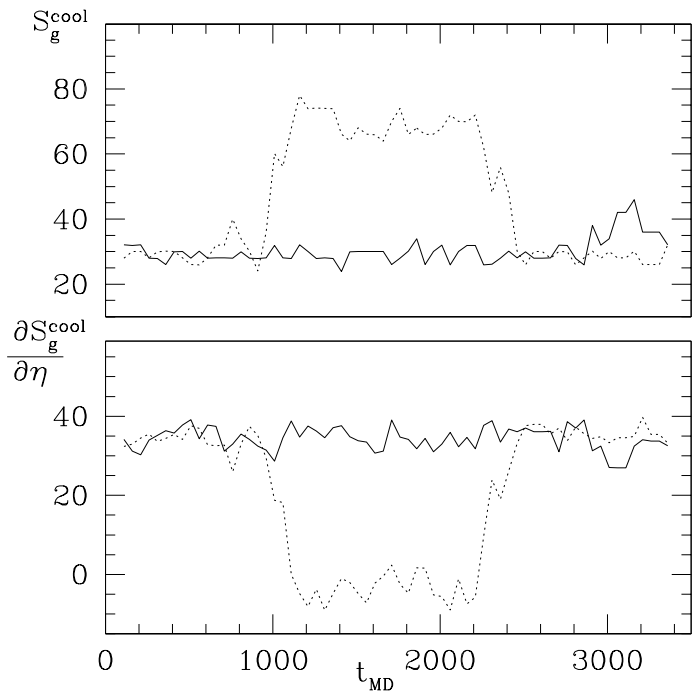,width=0.7\linewidth}
\vspace{-0.5cm}
    \caption{\fs Monte Carlo histories of $S_{\rm g}^{\rm cool}$ and
      $\partial S_{\rm g}^{\rm cool} / \partial \eta$ from two independent
      replica (solid and dotted lines) in a
      simulation of a $16^4$ lattice at $\gbar^2\approx 3.3$.
    }
    \label{fig_meta}
  \end{center}
\vspace{-0.5cm}
\end{figure}

2. For the two largest couplings discussed here the distribution of
$\partial S/\partial\eta$ shows long tails toward negative
values. The same effect was also observed in the computation of the
Schr\"odinger functional coupling in pure SU(3) gauge
theory~\cite{alpha:SU3}. 
We have related this tail to secondary local minima of the action
\cite{DellaMorte:2002vm} by measuring on cooled configurations the pure 
gauge contribution to the action $S_{\rm g}^{\rm cool}$ and to the coupling 
$\partial S_{\rm g}^{\rm cool} / \partial \eta$.
This leads to metastabilities as shown for
an $L/a=16$ simulation with $\gbar^2\approx 3.3$ in 
figure~\ref{fig_meta}. 
The upper panel is the
Monte Carlo history ($t_{\rm MD}$ is the Monte Carlo time in units of 
molecular dynamics trajectories) of the gauge part of the action after 
cooling for two independent replica. The lower panel shows the history of 
$\partial S_{\rm g}^{\rm cool} / \partial \eta$ for the same two replica.
The correlation between metastable states and small (even negative)
values of $\partial S_{\rm g}^{\rm cool} / \partial \eta$ appears evident in 
this case.
The action $S_{\rm g}^{\rm cool}$ for the metastability in the figure is
consistent with the value for a secondary solution of the field 
equations~\cite{alpha:SU3}, given our choice for the boundary fields. 
Numerical evidence suggests that this solution is a local minimum.

In order to estimate the weight of these contributions in our 
expectation values properly,
we have enhanced their occurrence through a modified sampling
similar to~\cite{alpha:SU3},
adding to the HMC effective action a term
\begin{equation}\label{eq:gammaterm}
\left.\gamma\,{{\partial S_{\rm g}}\over{\partial \eta}} \right|_{\eta=0} +
{{1}\over{w_{\gamma}}}\,(\gamma-\gamma_0)^2 \,,
\end{equation}
where $\gamma_0$ and $w_{\gamma}$ are fixed to suitable (positive) values, 
while $\gamma$ is a dynamical variable. The expectation values in the
original ensemble are then obtained by reweighting. 
By some deterministic cooling procedure with a fixed number of cooling 
steps we now define a quantity $q$ whose value is 1 for metastable 
configurations, and 0 otherwise, such that 
$\delta_{q1} + \delta_{q0} \equiv 1$.
For an arbitrary observable $O$ we have an exact identity
\begin{eqnarray}
\langle O\rangle &=& \langle \delta_{q1}O\rangle +
\langle\delta_{q0}O\rangle
\\ \nonumber
&=&\langle \delta_{q1}O\rangle + \langle O \rangle_{\bar{1}}
(1-\langle \delta_{q1}\rangle) \, ,
\end{eqnarray}
with $\langle O \rangle_{\bar{1}}=\langle \delta_{q0}O\rangle/
\langle \delta_{q0}\rangle$. If the main contribution to
$\langle O \rangle$ comes from the configurations with $q=0$,
a precise estimate of $\langle O \rangle$ just requires
a precise estimate of $\langle O \rangle_{\bar{1}}$, which can be
obtained by an  algorithm that samples only the
$q=0$ sector, together with  rough estimates  of
$\langle \delta_{q1}\rangle$
and  $\langle \delta_{q1}O\rangle$ that can be
obtained by the modified sampling.

At $\gbar^2\approx 3.3$ we get $\langle \delta_{q1}\rangle=0.3(2)\%$,
independent of $L/a=8,12$ within the error. The effect of metastable states
on the coupling is $0.10(2) \%$.  At the coupling $\gbar^2\approx 5.5$
the occurrence of metastable states is much more frequent, as expected, and we
get $\langle\delta_{q1}\rangle=5(1)\%$. At the same time their sampling is much
easier already in the original ensemble, using either the PHMC or the
HMC algorithm.   
In fact, we have repeated the $L/a=12$ simulation for 
$\bar{g}^2 \approx 5.5$ using PHMC for the ordinary ensemble 
(without (\ref{eq:gammaterm})) as in~\cite{alpha:letter}, but measuring 
in addition the occurrence of metastable states. 
This turned out to be around $6\%$, and for $\bar{g}^2$ we have obtained a 
result fully consistent with the number in~\cite{alpha:letter}.

In summary, we can be confident that topologically non-trivial
sectors are irrelevant for our observables with our
choice of parameters and at the present
level of precision. In contrast, there are secondary minima in
the action, which are visible as metastable states in
the Monte Carlo sequence. They are relevant starting at $\gbar^2\approx3$
and have been taken into account efficiently by deviating
from naive importance sampling and combining
two different properly chosen ensembles.

\section{Results}
\label{sec:results}

\subsection{The strong coupling}
\begin{figure}[htbp]
  \begin{center}
    \psfig{file=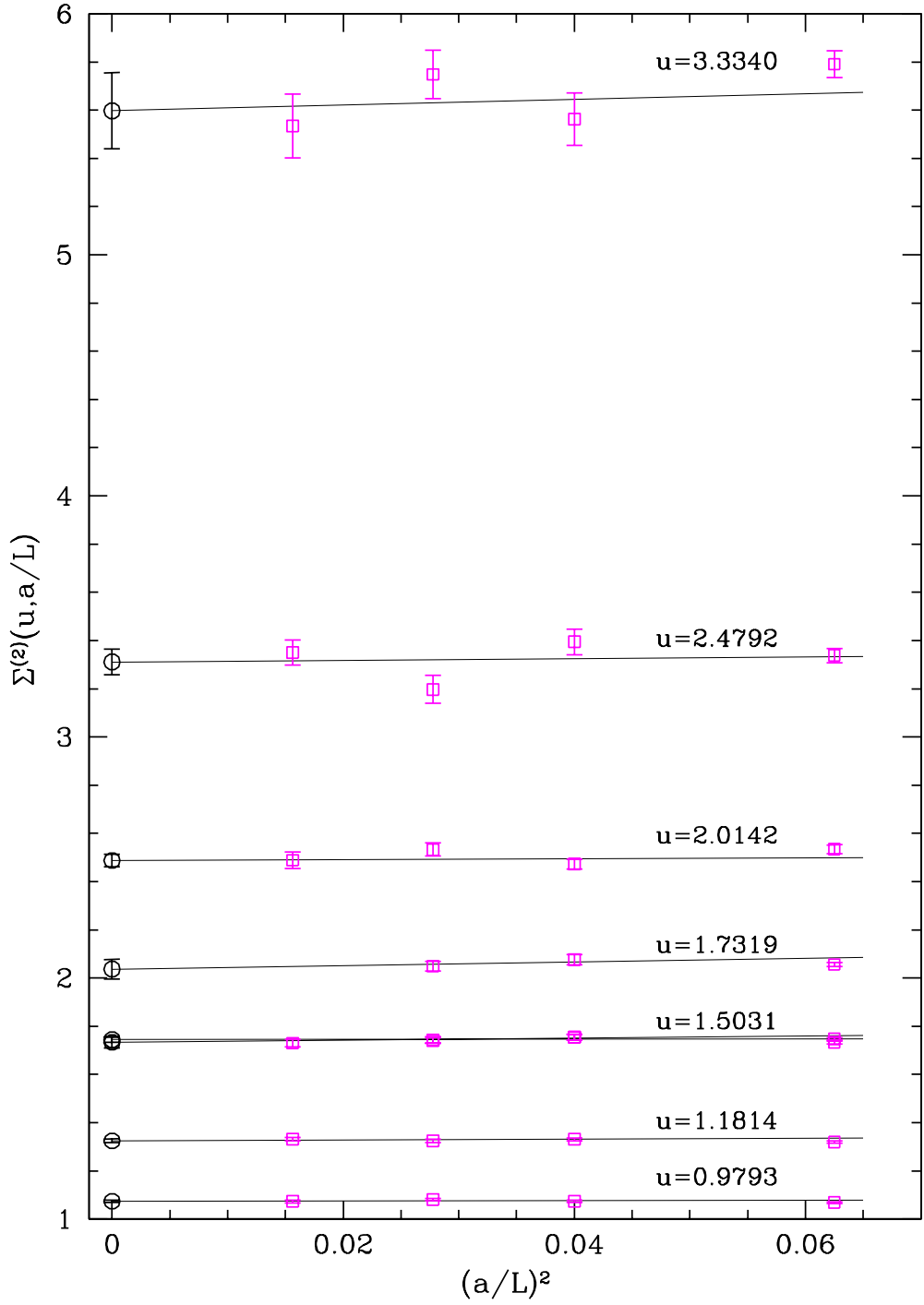,width=\linewidth}
\vspace{-0.75cm}
    \caption{\fs Continuum extrapolation of the step scaling function.}
    \label{fig:context1}
  \end{center}
\end{figure}

In figure~\ref{fig:context1} we show the approach of the
step scaling function $\Sigma^{(2)}(u,a/L)$ to the continuum limit. 
The cutoff effects are small. Actually, all the
data are compatible with constants. If we use simple fits
to constants, the combined $\chi^2$ per degree
of freedom for all the eight continuum extrapolations is about 1.4
regardless of the number of lattices included in the fit. Even the
points at $L/a=4$ are compatible with a constant continuum
extrapolation. One possible strategy for the continuum extrapolation
is thus a fit to a constant that uses the lattices with $L/a=6,8$.

In this fit to constants we exclude the two coarsest lattices since
there is always the danger of including systematic cutoff 
effects into the results coming from the lattices with large
$a/L$. Therefore, we have also investigated two alternative fit 
procedures. The first and most conservative one 
(denoted as ``global fit'' in the tables)
is a combined continuum
extrapolation of all the data sets, but excluding $L/a=4$.
Here we use the two-parameter ansatz
\begin{eqnarray}
\label{eq:rho}
\Sigma^{(2)}(u,a/L)= 
\sigma(u)+\rho^X\,u^4\,(a/L)^2\,,\quad 
X\in\{\text{1-loop},\text{2-loop}\}\,,
\end{eqnarray}
for the lattice artifacts, where the coefficients $\rho^X$ are understood
to be associated to the data with the 1-loop and the 2-loop value of $\ct$,
respectively.\footnote{We have also considered fits that use a
common ansatz for the step
scaling function for all the data sets. These lead to slightly smaller
error bars of the final results.} 
This fit results in
$  \rho^{\text{1-loop}} = 0.08(13)$ and 
$  \rho^{\text{2-loop}} = 0.01(4)$,
quantifying that lattice artifacts are not detectable in our data. 
Moreover we have studied a mixed fit procedure, using a fit to 
constants for the lattices with 
$L/a=6,8$ for the 2-loop improved data sets and 
the global fit ansatz~(\ref{eq:rho}), which includes a slope for the cutoff 
effects, for the 1-loop improved data sets. 

These different fits are 
performed to investigate the uncertainties in the continuum results.
All our plots refer to the most conservative of these three fit
procedures, which leaves both $\rho^{\text{1-loop}}$ and $\rho^{\text{2-loop}}$ 
unconstrained.

\begin{table}[ht]
  \centering
  \begin{tabular}{llll}\hline\\[-1.5ex]
    $u$ & \multicolumn{3}{c}{$\sigma(u)$} \\ 
        & global fit  & fit to constants & mixed fit \\[0.5ex] \hline\\[-1.5ex]
0.9793  & 1.0736(44)  & 1.0778(36)  & 1.0736(44) \\ 
1.1814  & 1.3246(81)  & 1.3285(50)  & 1.3246(81) \\ 
1.5031  & 1.733(23)   & 1.741(12)   & 1.733(23)  \\ 
1.7319  & 2.037(41)   & 2.049(20)   & 2.037(41)  \\[0.5ex] \hline\\[-1.5ex]
1.5031  & 1.7440(97)  & 1.738(10)   & 1.738(10)  \\ 
2.0142  & 2.488(26)   & 2.516(22)   & 2.516(22)  \\ 
2.4792  & 3.311(52)   & 3.281(39)   & 3.281(39)  \\ 
3.3340  & 5.60(16)    & 5.670(80)   & 5.670(80)  \\[0.5ex]
\hline
  \end{tabular}
  \caption{\fs Continuum limit of the step scaling function.}
  \label{tab:ssf}
\end{table}
The results of the continuum limit extrapolation of the step scaling 
function $\Sigma^{(2)}(u,a/L)$ are recorded in table~\ref{tab:ssf}.
At $u=1.5031$ we have two sets of data, one of which was produced with the
1-loop value and the other with the 2-loop value of $\ct$.
Both continuum results agree well within their errors, which is an 
independent check of our extrapolation procedures.

\begin{figure}[htbp]
  \begin{center}
\vspace{-2.3cm}
      \psfig{file=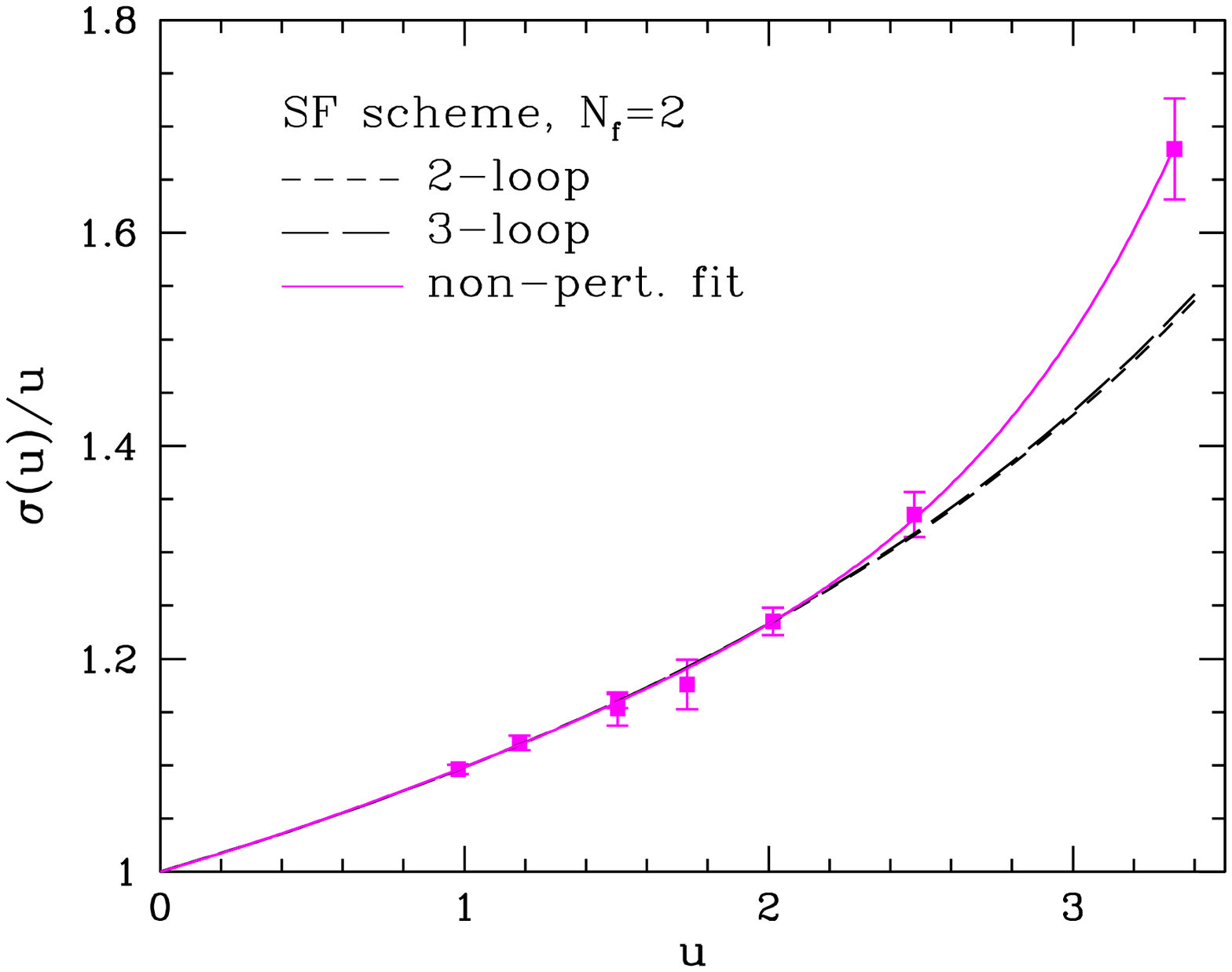,
      width=0.7\linewidth}
\vspace{-1cm}
    \caption{\fs Step scaling function $\sigma(u)$. The dashed lines show
      the perturbative results from the integration of the 2-loop and 
      3-loop $\beta$-function, respectively.}
    \label{fig:sigma}
  \end{center}
\end{figure}

We interpolate the values of table~\ref{tab:ssf} by a polynomial of
degree~6 in $u$, the first coefficients up to $u^3$ being fixed by
2-loop perturbation theory, cf.~(\ref{eq:sigmaPT}). This
interpolation is depicted in figure~\ref{fig:sigma}.
For small values of $u<2$, the step scaling function is well described
by perturbation theory. Actually, the perturbative lines
shown in this plot are the solution of
\begin{equation}
  -2\ln 2 = \int_u^{\sigma(u)}\!\frac{dx}{\sqrt{x}\ \beta(\sqrt{x})}
\end{equation}
for $\sigma(u)$ using the 2-loop and the 3-loop
$\beta$-function. 
Looking at the comparison of successive perturbative approximations
to the non-perturbative results, it appears likely that higher orders
would not improve the agreement at the largest coupling. Rather we
appear to have reached a value for the coupling, where the perturbative
expansion has broken down. In fact, already in section~\ref{sec:algo}
we have discussed the indications that fluctuations around a
secondary minimum of the action are important at this value
of the coupling. Such a mechanism may represent one {\em possible}
source of non-perturbative effects.\footnote{In this context we note
further that for the pure gauge theory it has been demonstrated that the
Schr\"odinger functional coupling grows exponentially with $L$ at even
larger values of $L$ \cite{lat01:jochen}.
We do not expect that any semiclassical picture is applicable in
that regime but rather see this as a disorder phenomenon.}

We adopt the parametrized form of the step scaling function to compute
the $\text{$\Lambda$-parameter}$. To this end we start a recursion with a
maximal coupling $u_\text{max} = \gbar^2(L_\text{max})$. 
For $u_{\rm max}=5.5$, the recursive step~(\ref{eq:rec})
     is solved numerically to get the couplings $u_i$,
     corresponding to the energy scales $\mu = 2^i/L_{\rm max}$,
     that are quoted in table~\ref{tab:rec}. We then insert these couplings
     into eq.~(\ref{eq:lambda}) for the $\Lambda$-parameter, using there the
     3-loop $\beta$-function.
\begin{table}[htbp]
  \centering
  \begin{tabular}{lllllll}\hline\\[-1.5ex]
    \multicolumn{3}{c}{global fit} &
    \multicolumn{2}{c}{constant fit, $L/a=6,8$} &
    \multicolumn{2}{c}{mixed cont.~extrap.} \\
    $i$ & $u_i$ & $-\ln(\Lambda L_\text{max})$ & $u_i$ & $-\ln(\Lambda L_\text{max})$& $u_i$ & $-\ln(\Lambda L_\text{max})$\\[0.5ex] \hline\\[-1.5ex]
 0    &  5.5        &    0.957            &  5.5         &    0.957     &   5.5         &    0.957       \\
 1    &  3.306(40)  &    1.071(25)        &  3.291(18)   &    1.081(12) &   3.291(19)   &    1.081(12)   \\
 2    &  2.482(31)  &    1.093(37)        &  2.479(20)   &    1.096(23) &   2.471(20)   &    1.106(24)   \\
 3    &  2.010(27)  &    1.093(48)        &  2.009(19)   &    1.096(35) &   2.003(19)   &    1.106(35)   \\
 4    &  1.695(22)  &    1.089(57)        &  1.691(16)   &    1.099(43) &   1.690(17)   &    1.103(44)   \\
 5    &  1.468(18)  &    1.087(65)        &  1.462(14)   &    1.109(49) &   1.464(15)   &    1.100(52)   \\
 6    &  1.296(16)  &    1.086(73)        &  1.288(12)   &    1.122(55) &   1.292(14)   &    1.100(63)   \\
 7    &  1.160(14)  &    1.086(82)        &  1.151(11)   &    1.138(62) &   1.157(13)   &    1.101(74)   \\
 8    &  1.050(13)  &    1.088(93)        &  1.041(10)   &    1.155(70) &   1.048(13)   &    1.103(87)   \\[0.5ex] \hline
  \end{tabular}
  \caption{\fs Recursive computation of the $\Lambda$-parameter starting
 at $u_0=u_\text{max}=5.5$.}
  \label{tab:rec}
\end{table}
This gives the results in the third column of
table~\ref{tab:rec}. Employing the 2-loop $\beta$-function leads to
results that are larger by roughly $0.02$. The table shows that for
$u<2$ the $\Lambda$-parameter barely moves within its error bars. To be
conservative, we use the global fit result and quote 
\begin{equation}
  -\ln(\Lambda L_\text{max}) = 1.09(7) \quad\text{at}\quad u_\text{max}=5.5
\end{equation}
as our final result, if the hadronic scale $L_\text{max}$ is defined
through $u_\text{max}=5.5$. 

We have in addition computed the $\Lambda$-parameter as a function of
$u_\text{max}$ in the interval $u_\text{max}=3.0\ldots 5.5$. The
results can be parametrized as
\begin{equation}
  \label{eq:lll}
  -\ln(\Lambda L_\text{max}) = \frac{1}{2b_0\,u_\text{max}} +
  \frac{b_1}{2b_0^2} \ln(b_0\,u_\text{max}) - 0.1612 + 0.0379\,u_\text{max}\,.
\end{equation}
This parametrization is motivated by~(\ref{eq:lambda}) and represents
the central values of our
data with a precision better than one permille. The absolute
error of $-\ln(\Lambda L_\text{max})$
that we quote for all the values of $u_\text{max}$ is $0.07$. This means
that we have calculated the $\Lambda$-parameter in units of $L_\text{max}$
with a precision of seven percent.

\begin{figure}[htbp]
  \begin{center}
\vspace{-2cm}
      \psfig{file=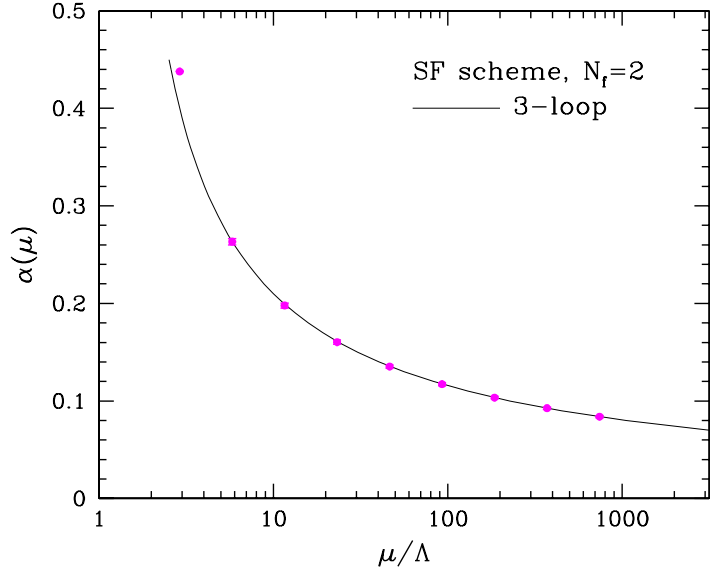,
      width=0.7\linewidth}
\vspace{-1cm}
    \caption{\fs Running of the strong coupling in the Schr\"odinger
      functional scheme. }
    \label{fig:running}
  \end{center}
\end{figure}

The running of the Schr\"odinger functional coupling 
$\alpha(\mu)=\gbar^2(1/\mu)/(4\pi)$ as a function of $\mu/\Lambda$ is 
displayed in figure~\ref{fig:running}. 
\begin{figure}[htbp]
  \begin{center}
\vspace{-2cm}
      \psfig{file=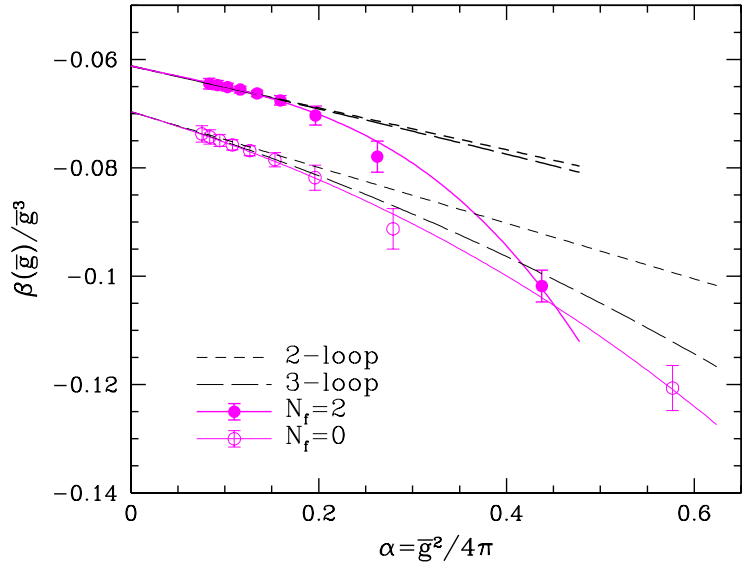,
      width=0.7\linewidth}
\vspace{-1cm}
    \caption{\fs Non-perturbative $\beta$-function in the Schr\"odinger
      functional scheme. }
    \label{fig:beta}
  \end{center}
\end{figure}
The points refer to the entries of the second column of
table~\ref{tab:rec}. The symbol size is larger than their
error. The difference between the perturbative and the 
non-perturbative running of the coupling looks small in this
plot. However, if we had used perturbation theory only to evolve the
coupling over the range considered here, 
the $\Lambda$-parameter would have been overestimated by up
to $14\%$, depending on $u_\text{max}$. This corresponds to an extra
error of $3\%$ for the coupling in the range where its value is close
to 0.12, corresponding to the physical value of $\alpha_\MSbar$ at
$M_{\rm Z}$. Needless to add, this error could of course not even be
quantified without non-perturbative information.

Furthermore, our non-perturbative coupling was designed
to have a good perturbative expansion, since we rely on the 3-loop
$\beta$-function in the (high-energy part of the) computation of the 
$\Lambda$-parameter. 
With our computation we have {\em shown} that for the Schr\"odinger
functional coupling there is an overlapping
region where both, perturbative and non-perturbative methods apply. In
no way is this to be interpreted as a general statement about QCD
observables or couplings at certain energies.

In figure~\ref{fig:beta}
we show the non-perturbative $\beta$-function in the Schr\"odinger
functional scheme, together with the 2-loop and the 3-loop
perturbation theory. It has been obtained recursively
from~(\ref{eq:betarec}). The derivative of the step scaling 
function needed there has been calculated from the polynomial
interpolating $\sigma(u)$ (see continuous line in figure~\ref{fig:sigma}).
The non-perturbative data are fitted with two parameters
beyond the 2-loop $\beta$-function.
The plot again shows an overlapping region in $\gbar$,
where the perturbative and the non-perturbative $\beta$-functions
agree well with each other. For $\alpha>0.2$, however, perturbation
theory is no longer valid. Furthermore, the plot shows the difference
between $\Nf=0$ and $\Nf=2$. Already the leading coefficient $b_0$ of
the $\beta$-function depends on the number of flavours, and this is
nicely reflected in the figure.

\subsection{Computation of $\vbar$ as a function of the strong coupling}

The difference between the quenched approximation and the two-flavour
theory is also apparent in the renormalized
quantity $\vbar$ defined in~(\ref{eq:vbar}). As a function of the
coupling $u$ we write (at zero quark mass)
\begin{equation}
  \vbar = \omega(u) = \lim_{a/L\rightarrow 0} \Omega(u,a/L)\,.
\end{equation}
In perturbation theory, $\Omega$ is known to 2-loop order,
\begin{equation}
  \Omega(u,a/L) = (v_1 + v_2 u) \left( 1 + \epsilon_1(a/L) +
  \epsilon_2(a/L) u \right) + \rmO(u^2)\,.
\end{equation}
Here~\cite{alpha:SU3,pert:1loop,Bodeunpub},
\begin{eqnarray}
  v_1 &=& 0.0694603(1) + 0.0245370(1) \Nf\,, \\
  v_2 &=& -0.001364(14)-0.000101(17) \Nf - 0.0003362(30) \Nf^2\,,
\end{eqnarray}
and the perturbative cutoff effects are listed in
table~\ref{tab:epsilon}.
Note that the tree-level coefficient of $\vbar$ vanishes exactly
because of the definition of the couplings. 
The perturbative results indicate a large effect for going from the
zero- to the two-flavour theory. 
\begin{table}[htbp]
  \vspace{0.25cm}
  \centering
  \begin{tabular}[c]{ccccccr@{.}l}\hline\\[-1.5ex]
    && \multicolumn{2}{c}{$\Nf=0$} && \multicolumn{3}{c}{$\Nf=2$} \\
    $L/a$ && $\epsilon_1(a/L)$ & $\epsilon_2(a/L)$ && $\epsilon_1(a/L)$
    & \multicolumn{2}{c}{$\epsilon_2(a/L)$} \\[0.5ex] \hline\\[-1.5ex]
    4 && 0.1880609(17) & $-0.02020(24)$ && 0.3044344(25) & $0$&$03725(43)$ \\
    5 && 0.1085045(16) & $-0.01347(22)$ && 0.1822083(22) & $0$&$01255(39)$ \\
    6 && 0.0677292(15) & $-0.00910(22)$ && 0.1114033(21) & $0$&$00325(36)$ \\
    7 && 0.0460993(15) & $-0.00660(21)$ && 0.0727865(20) & $-0$&$00047(35)$ \\
    8 && 0.0335967(15) & $-0.00513(21)$ && 0.0502340(20) & $-0$&$00130(34)$ \\
    10&& 0.0203927(15) & $-0.00357(21)$ && 0.0280832(19) & $-0$&$00118(34)$ \\
    12&& 0.0138002(15) & $-0.00273(20)$ && 0.0181305(19) & $-0$&$00091(33)$ \\
    14&& 0.0099904(15) & $-0.00219(20)$ && 0.0127946(19) & $-0$&$00074(33)$ \\
    16&& 0.0075781(15) & $-0.00181(20)$ && 0.0095635(19) & $-0$&$00064(33)$ \\
    20&& 0.0047983(14) & $-0.00133(20)$ && 0.0059634(19) & $-0$&$00050(33)$ \\
    24&& 0.0033132(14) & $-0.00103(20)$ && 0.0040873(19) & $-0$&$00040(33)$ 
    \\[0.5ex] \hline
  \end{tabular}
  \caption{\fs Cutoff effects of $\vbar$ in perturbation theory.}
  \label{tab:epsilon}
\end{table}

The first step in the analysis is to project $\vbar$ on zero mass. To
this end we have obtained the crude estimate 
$\partial \vbar /\partial (am) \approx -0.15(4)$ at constant $u$ from the 
matching runs at the smallest coupling and with $L/a=4$. 
We use it also at the other couplings. After this projection, the
perturbative cutoff effects are eliminated (similarly
to~(\ref{eq:Sigma2})) by replacing $\Omega$ by
\begin{equation}
  \Omega^{(2)}(u,a/L) = \frac{\Omega(u,a/L)}{1 + \epsilon_1(a/L) +
  \epsilon_2(a/L) u}
\end{equation}
in the analysis. In contrast to the coupling, this correction is
substantial and the resulting continuum extrapolation is much
smoother~\cite{alpha:letter}. 

Then we project $\Omega^{(2)}(u,a/L)$ on some reference couplings in
the range $u=0.9793 \ldots 5.5$, using a numerical estimate for the slope. 
In principle, we would have to 
propagate the error of $u=\gbar^2$ into an extra error of
$\Omega^{(2)}(u,a/L)$. However, it turns out, both from perturbation
theory and from the non-perturbative fits later, that the variation of
$\vbar$ with $u$ is so small that this extra error can safely be
neglected. 

We make an ansatz linear in $(a/L)^2$ for the continuum
extrapolation. The data for the two different
actions are extrapolated separately and the continuum results are then
averaged according to their weight. The data at $L/a=4,5$ are left
out. 

\begin{figure}[htbp]
  \begin{center}
\vspace{-2.275cm}
      \psfig{file=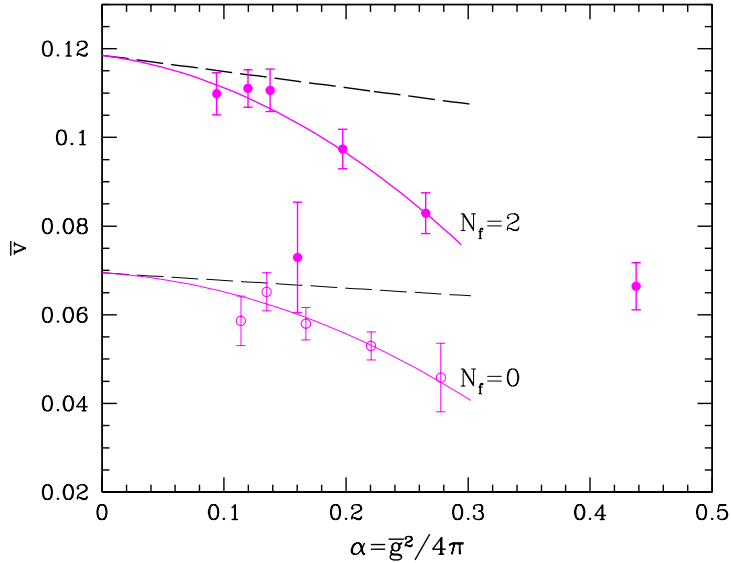,
      width=0.7\linewidth}
\vspace{-1cm}
    \caption{\fs $\vbar$ as a function of the strong coupling
      $\alpha$. The dashed lines show the 2-loop perturbation
      theory while the solid lines are non-perturbative fits
      for $\alpha < 0.3$ including an additional $\alpha^2$-term.}
    \label{fig:vbar}
  \end{center}
\end{figure}

The results are displayed in figure~\ref{fig:vbar}
together with $\vbar$ at $\Nf=0$ from ref.~\cite{alpha:SU3}. 
The comparison shows that $\vbar$ increases by almost a factor two when 
going from the quenched approximation to $\Nf=2$. 

\subsection{Evaluation of $\Lambda r_0$ \label{sect:lambda_r0}}

For the O($a$) improved action~\cite{impr:csw_nf2} used in our computations,
the low-energy scale $r_0$ has been calculated at $\beta\equiv6/g_0^2=5.2$
(or $a\approx0.1\,$fm) by two groups~\cite{UKQCD:nf2b52,JLQCD:nf2b52}. 
Recently, these large-volume $\nf=2$ simulations of QCD have been extended 
to smaller values of the lattice spacing, namely 
$5.2 \leq \beta \leq 5.4$~\cite{QCDSF:nf2mstrange}. 
In order to obtain $\Lambda r_0$ and to study its dependence on the lattice 
spacing, we here use results for $r_0/a$ of~\cite{QCDSF:nf2mstrange} at 
$\beta=5.2,5.29,5.4$ and compare also to the numbers resulting from $r_0/a$ 
of~\cite{JLQCD:nf2b52}.

\begin{table}[htb]
\begin{center}
\begin{tabular}{cccll}
\hline \\[-1.5ex]
 & & &  $c_{\rm t}$=1-loop & $c_{\rm t}$=2-loop \\
  $\beta$ & $\kappa$ &  $L/a$    &  \multicolumn{2}{c}{$\gbar^2(L)$} \\[0.5ex]
\hline\\[-1.5ex]
 5.20 & 0.13600 & 4 & 3.32(2) & 3.65(3)\\
 5.20 & 0.13600 & 6 & 4.31(4) & 4.61(4)\\[1ex]
 5.29 & 0.13641 & 4 & 3.184(16) & 3.394(17)\\
 5.29 & 0.13641 & 6 & 4.059(32) & 4.279(37)\\
 5.29 & 0.13641 & 8 & 5.34(8)   & 5.65(9)\\[1ex]
 5.40 & 0.13669 & 4 & 3.016(20)& 3.188(24)\\
 5.40 & 0.13669 & 6 & 3.708(31)& 3.861(34)\\
 5.40 & 0.13669 & 8 & 4.704(59)& 4.747(63)\\[1ex]
\hline
\end{tabular}
\caption{\fs Simulation results for $\gbar^2(L)$ at low $\beta$. The hopping
	parameters $\kappa$ are set to the critical ones ($\kappa_{\rm c}$) 
        of \protect\cite{QCDSF:nf2mstrange}.}
\label{gsq_lowbeta_tab}
\end{center}
\end{table}

First, we obtain the renormalized coupling on lattices
with extent $L/a=4,6,8$ at the three chosen values of $\beta$.
It is listed in table~\ref{gsq_lowbeta_tab}.
The hopping parameters $\kappa$ are taken from \cite{QCDSF:nf2mstrange}.
They correspond
to roughly massless pions and thus massless quarks. We checked 
that reasonable changes of $\kappa$, e.g. 
requiring  eq.~(\ref{eq:massenull}), affect our analysis only
to a negligible amount. We then
set the improvement coefficient $c_{\rm t}$
to its 2-loop value and
obtain $L_{\rm max}/a$ for the three values of
$\beta$ combined with two fixed values 
$u_{\rm max}\equiv\gbar^2(L_{\rm max}) = 3.65$ and 
$u_{\rm max} = 4.61$ by an interpolation of
the data in table~\ref{gsq_lowbeta_tab}. 
These values of $L_{\rm max}/a$, which are recorded in table~\ref{match_tab},
are insensitive to the interpolation formula used.

\begin{table}[htb]
\begin{center}
\begin{tabular}{cccccc}
\hline \\[-1.5ex]
         &        & \multicolumn{2}{c}{$u_{\rm max}=3.65$} 
                  & \multicolumn{2}{c}{$u_{\rm max}=4.61$} \\[0.5ex]
 $\beta$ & $r_0/a$ & $L_{\rm max}/a$ & $\Lambda_\MSbar\, r_0$ &
                        $L_{\rm max}/a$ & $\Lambda_\MSbar\, r_0$ \\[0.5ex]
\hline\\[-1.5ex]
 5.20 &  5.45(5)(20) &  4.00(6) & 0.655(27) &  6.00(8)  & 0.610(25) \\           
 5.29 &  6.01(4)(22) &  4.67(6) & 0.619(25) &  6.57(6)  & 0.614(24) \\         
 5.40 &  7.01(5)(15) &  5.43(9) & 0.621(17) &  7.73(10) & 0.609(16) \\[0.5ex]     
\hline
\end{tabular}
\caption{\fs Low-energy scale $r_0$ in the chiral limit and the 
	combination $\Lambda_\MSbar\, r_0$ as obtained for
	two values of $u_{\rm max}=\gbar^2(L_{\rm max})$.
\label{match_tab}}
\end{center}
\end{table}

\begin{figure}[htbp]
  \begin{center}
      \psfig{file=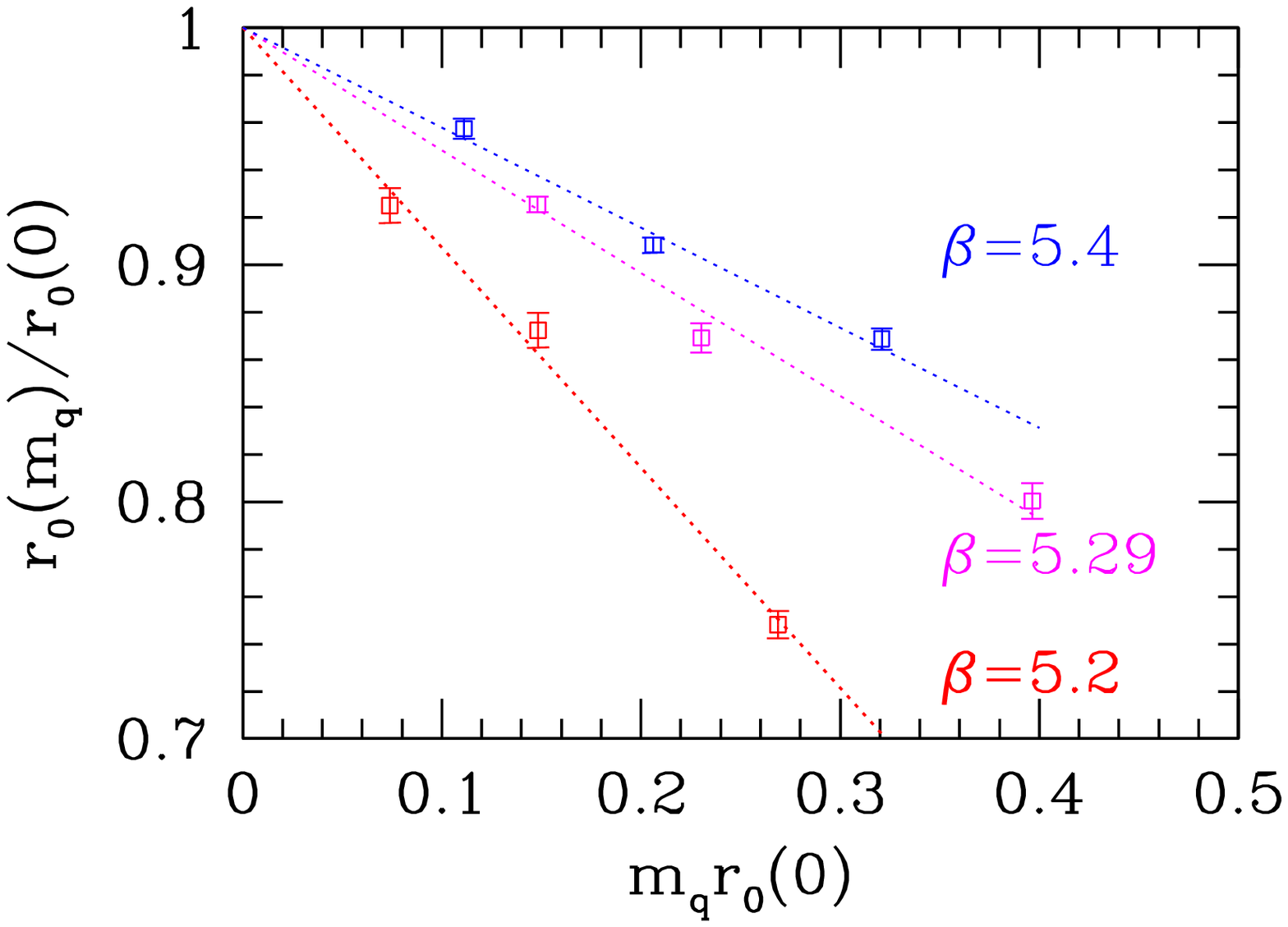,width=0.60\linewidth}
\vspace{-0.25cm}
    \caption{\fs Dependence of $r_0$ on the bare quark mass 
        $m_{\rm q}=(1/\kappa - 1/\kappa_{\rm c})/2$. Both quantities
        are rescaled (made dimensionless) by the extrapolated value of 
        $r_0$, denoted by $r_0(0)$. The uncertainty of this
        rescaling is not propagated into the errors. For $\kappa_c$,
        we take the values of  \protect\cite{QCDSF:nf2mstrange}, listed in
        table~\ref{gsq_lowbeta_tab}.}
    \label{fig:r0mq}
  \end{center}
\end{figure}

Second, we analyze the raw data for $r_0/a$ at finite bare 
quark masses, $m_{\rm q}$,
in order to obtain the value corresponding to massless quarks
(the up- and down-quark masses may safely be neglected in
this context). 
At each bare coupling, three different quark
masses, corresponding to pion masses between about
$500\,\MeV$ and $1\,\GeV$, have been simulated in \cite{QCDSF:nf2mstrange}. 
As seen in figure~\ref{fig:r0mq},  the radius $r_0$ depends
approximately linearly on the bare quark mass in this range. 
The figure also demonstrates that
the slope is strongly cutoff dependent; 
its magnitude decreases quite rapidly as $\beta$ increases
(the lattice spacing decreases).\footnote{
        The slope visible in figure~\ref{fig:r0mq} is
        not directly a physical observable, since (up to $a$-effects)
        the renormalized quark mass is given by
        $\mr = \zm \mq$ rather than by $\mq$. However, 
        it appears unlikely that the 
        strong dependence of the slope on $g_0$ is cancelled
        by $\zm$, since
        the latter is expected to be a weak function
        of the bare coupling. More details can be found
        in \cite{lat03:rainer}, where also a properly renormalized
        slope has been analyzed.}
This has been noted earlier~\cite{lat03:rainer}
and reminds us that the study of lattice artifacts 
is important. On the other hand, we are here not
interested in the slope but in $r_0$ at zero quark mass. 
We estimate it by a simple linear extrapolation in $m_{\rm q}$. 
Since the linear behaviour is not guaranteed to extend
all the way to zero quark mass, we
include a systematic error of the extrapolation in addition to the 
statistical one: an uncertainty of half the difference
of the last data point and the extrapolated value
is added as the second error in table~\ref{match_tab}.  
Within the total error, dominated by the systematic
one due to the extrapolation, our values of $r_0/a$ do agree with those 
quoted in
\cite{UKQCD:nf2b52,JLQCD:nf2b52,QCDSF:nf2mstrange},
where somewhat different ans\"atze -- and in \cite{JLQCD:nf2b52}
also different data -- have been used. 

Third, combining eqs.~(\ref{eq:lll}) and (\ref{eq:lambdamsbar})
with $r_0/a$ and $L_{\rm max}/a$ of  table~\ref{match_tab},
one arrives at the columns $\Lambda_\MSbar\, r_0$ in that table.
Here we have added errors in quadrature,
except for the one attributed to  eq.~(\ref{eq:lll}), which
is independent of the bare coupling. The resulting 
$\Lambda_\MSbar\, r_0$ are remarkably stable with respect
to the change of the lattice spacing and also the
choice of $u_{\rm max}$. In particular, 
for $L_{\rm max}/a > 4.5$ all numbers including their 
errors are covered by the interval
$0.58 \leq \Lambda_\MSbar\, r_0 \leq 0.66$ and also the
central value obtained with the worst discretization 
$L_{\rm max}/a=4$ is inside.
We thus quote 
\begin{equation}
  \label{eq:lamr0}
  \Lambda_\MSbar^{(2)} \, r_0 = 0.62(4)(4)
\end{equation}
as our result, the second error deriving from the $7\%$ error 
on $\Lambda L_{\rm max}$. 

We finally mention that we repeated the above analysis 
also for $c_{\rm t}$ to 1-loop precision and 
the values $u_{\rm max} = 3.32$ and $4.31$ suggested
by table~\ref{gsq_lowbeta_tab}. 
The central values for $\Lambda_\MSbar\, r_0$ are then up to
$15\%$ lower than for $c_{\rm t}$ at 2-loop precision,
but this difference shrinks as $L_{\rm max}/a$ grows. 
At the largest value of $L_{\rm max}/a$, the
number for  $\Lambda_\MSbar\, r_0$ is again fully contained in the 
range $\Lambda_\MSbar \, r_0 = 0.62(4)$.


\section{Conclusions}

This non-perturbative QCD computation has required  
extensive simulations of $\nf=2$ QCD with O($a$) improved 
massless Wilson fermions in finite volume. In our situation,
discretization errors turned out to be very small, as seen for example in 
figure~\ref{fig:context1} and also in \cite{alpha:letter}. 
Although we could
achieve the necessary precision only on lattices up to a size $16^4$,  
the smallness of discretization errors allowed to
obtain the running of the QCD coupling in the continuum
limit and to good accuracy. As in the $\nf=0$ (pure gauge) theory, 
the energy dependence of the coupling in the Schr\"odinger functional
scheme is now known over more than two orders of magnitude in the energy 
scale. This is the main result of our investigation.

The $\nf$-dependence is best illustrated in figure~\ref{fig:beta},
where we also observe excellent agreement with 3-loop perturbation
theory for $\alpha \leq 0.2$, while for larger $\alpha$, the 
non-perturbative $\beta$-function breaks away rapidly from
the perturbative approximation. At $\alpha\approx0.3\ldots 0.4$,
a couple of additional higher-order perturbative terms 
with {\em coefficients of a reasonable size} would not be able to come 
close to the non-perturbative $\beta$-function.

To calibrate the overall energy scale, one fixes a large
enough value of the coupling to be in the low-energy 
region and relates the associated distance, $L_{\rm max}$,
to a non-perturbative, large-volume observable. 
For technical reasons, explained in section~\ref{sect:hadronic},
we have chosen the hadronic radius $r_0$, which has an
unambiguous definition in terms of the force $F(r)$ between static 
quarks, via $r_0^2 F(r_0) = 1.65$~\cite{pot:r0}.
This quantity has also been chosen in the computation
of the $\Lambda$-parameter for  $\nf=0$ \cite{mbar:pap1}. 
We compare $\Lambda r_0$ for various numbers of flavours in
table~\ref{lambda_tab}.
\begin{table}[htb]
\begin{center}
\begin{tabular}{clcc}
\hline\\[-1.5ex]
   $\nf$  &  $\Lambda_\msbar^{(\nf)}\,r_0$  &  reference & remarks \\[0.5ex]
\hline\\[-1.5ex]
   0      &  0.60(5) &  \cite{mbar:pap1} & \\
   2      &  0.62(4)(4) &  this work  \\
   4      &  0.57(8)    &  \cite{alpha:blum1,pert:moch1,vanNeerven:1991nn} & 
                DIS \@ NNLO \& $r_0=0.5\,\fm$\\
   4      &   0.74(10)  &  \cite{Bethke:2004uy} & world average 
                \& $r_0=0.5\,\fm$\\ 
   5      &   0.54(8)   &   \cite{Bethke:2004uy} & world average
                \& $r_0=0.5\,\fm$\\[0.5ex]
\hline
\end{tabular}
\caption{\fs The QCD $\Lambda$-parameter in units of $r_0$. 
        Non-perturbative, purely theoretical determinations
        for $\nf=0,2$ are compared to extractions of $\Lambda$
        from high-energy scattering experiments, using high-order 
        perturbation theory combined with the phenomenological estimate 
        $r_0\approx0.5\,\fm$ \protect\cite{pot:r0}. 
        }
\label{lambda_tab}
\end{center}
\end{table}

The last two entries in the table represent one and the
same world average by S.~Bethke of $\alpha$-measurements.
They are related by the {\em perturbative} matching of the 
effective theories with 
$\nf=4$ and $\nf=5$ massless quarks \cite{alpha:bernwetz}.  
While little can be said on the $\nf$-dependence of 
$\Lambda_\msbar^{(\nf)}\,r_0$ on general grounds, the  prediction that there
should be a significant drop from $\nf=4$ to $\nf=5$ depends 
only on perturbation theory at the scale of the b-quark mass 
and should thus be reliable. A similar statement for the 
change from $\nf=3$ to $\nf=4$ is less certain as it involves physics at 
and below the mass of the charm quark. 
 
Another relevant issue in the above comparison is that we 
used  $r_0\approx0.5\,\fm$ to relate the high-energy experiments
to our theoretical predictions. Although it appears unlikely 
that $r_0$ differs by 10\% from this value, a true error is
difficult to estimate until a reliable
non-perturbative computation of e.g.~$r_0\fpi$ has been performed.
Indeed, such a computation, or more directly the computation
of $L_{\rm max} \times \fpi$ for $\nf=2$, is the most urgent next
step to be taken in our programme. After that, the effect of 
the remaining (massive) quarks needs to be estimated.    
 
Keeping the above caveats in mind, we still may convert the
$\Lambda$-parameter to physical units and obtain
\begin{equation}
   \Lambda_\MSbar^{(2)} = 245(16)(16) \,\MeV \quad 
   \text{[with $r_0=0.5\,\fm$]}\,.
\end{equation} 
Although in this case the four-flavour theory has not yet been reached,
it is a very non-trivial test of QCD that the non-perturbative
results, which use experimental input at low
energies of order $1/r_0\approx 400\,\MeV$, agree roughly with 
the high-energy, perturbative extractions of $\Lambda$. 
Unravelling the details in this comparison will still require   
some work; some of it was just mentioned.

Now, that $\alpha(\mu)$ is known, the tables presented in this work also 
provide the bare parameters of our lattice action needed 
in the computation of the energy dependence of the renormalized quark
mass and composite operators. These are then readily related to the 
appropriate renormalization group invariants. 

\section*{Acknowledgements}

We are indebted to Martin L\"uscher who founded the 
ALPHA Collaboration and who led ground-breaking work for this project
-- as demonstrated by the references we quote.
We further thank Achim Bode, Bernd Gehrmann, Martin Hasenbusch, 
Karl Jansen, Francesco Knechtli, Stefan Kurth, Hubert Simma, Stefan Sint, 
Peter Weisz and Hartmut Wittig for many useful discussions and 
collaboration in early parts of this project \cite{alpha:letter}. 
We are grateful to Gerrit Schierholz for communicating
the results of \cite{QCDSF:nf2mstrange}.
We further thank DESY for computing resources and the APE Collaboration 
and the staff of the computer centre at DESY Zeuthen for their support.

The computation of $\alpha_{\rm s}$ is one project of SFB Transregio 9
``Computational Particle Physics'' and has been strongly supported
there as well as in Graduierten\-kol\-leg GK 271 by the
Deutsche Forschungsgemeinschaft (DFG). We thank our colleagues
in the SFB for discussions, in particular 
Johannes Bl\"umlein, Kostia Chetyrkin and Fred Jegerlehner.
This work has also been supported by the
European Community's Human Potential Programme
under contract HPRN-CT-2000-00145.

\begin{appendix}

\section{Evaluation of $\partial \Gamma/\partial \eta$ \label{sect:dsdeta}}
Our central observable, eq.(\ref{eq_Gammap}), translates 
into the expectation value
\begin{equation}
\frac{\partial \Gamma}{\partial\eta}\bigg\vert_{\eta=0} = 
\left\langle \frac{{\rm d} S}{{\rm d} \eta} \right\rangle
 = \left\langle \frac{{\rm d} S_{\rm g}}{{\rm d} \eta} \right\rangle 
 + \left\langle \frac{{\rm d} S_{\rm f}^{\rm eff}}{{\rm d} \eta} \right\rangle\,,
\end{equation}
where the pure gauge part ${\rm d} S_{\rm g}/{\rm d} \eta$
has been discussed in \cite{alpha:SU3} and 
\begin{equation}
\label{eq_gbarferm1}
\frac{{\rm d} S_{\rm f}^{\rm eff}}{{\rm d} \eta}
= -\nf {\rm Tr}\, Q^{-1} \frac{{\rm d} Q}{{\rm d} \eta} 
= -\nf \,\left\langle\, \varphi^\dagger Q^{-1} \frac{{\rm d} Q}{{\rm d} \eta}
      \varphi \,\right\rangle_{\varphi} \,.
\end{equation} 
Here we have used $S_{\rm f}^{\rm eff} = -\nf  {\rm Tr} \ln Q$ with 
$Q=\gamma_5 (D+m_0) = Q^\dagger$ and $D$ the (one-flavour) 
Dirac operator including improvement terms. 
As usual, $S_{\rm f}^{\rm eff}$ is obtained after integrating out the 
fermion fields and the trace extends over colour and Dirac indices as well
as over the space-time points. The last expression
in eq.~(\ref{eq_gbarferm1}) represents an average over a 
complex random field $\varphi(x)$ with the property
\begin{equation}
\label{eq_gbarferm2}
   \langle \varphi^\ast_{c\alpha}(x)   \varphi_{d\beta}(y) 
     \rangle_{\varphi} = \delta_{cd}\,\delta_{\alpha\beta}\,\delta_{xy}\,,
\end{equation} 
where $c,d$ denote colour indices and $\alpha,\beta$ are Dirac indices.
We may finally rewrite ${\rm d} S_{\rm f}^{\rm eff}/{\rm d} \eta$ in the form 
\begin{eqnarray}
\label{eq_gbarferm3}
\frac{{\rm d} S_{\rm f}^{\rm eff}}{{\rm d} \eta}
& = &
-\nf \,\left\langle\, \sum_{x | x_0 \in \{a,T-a\}} \frac12 \tr\,
\frac{{\rm d} Q(x,x)}{{\rm d} \eta}
\left[\varphi(x) X^\dagger(x) + X(x) \varphi^\dagger(x)
\right]\right\rangle_{\varphi} \,,\nonumber\\
&   & 
Q X = \varphi \,,
\end{eqnarray} 
where we have used the fact that ${\rm d} Q/{\rm d} \eta$ 
vanishes except for the clover terms, which are diagonal in the coordinate
$x$ and contribute only on the time slices $x_0=a$ and $x_0=T-a$. 
Now ``tr'' is over spin and colour only.
Eq.~(\ref{eq_gbarferm3}) is in the form used in \cite{Jansen:1997yt} 
for the contribution of the clover terms to the pseudofermionic force 
in the HMC algorithm and is evaluated analogously. Only one solution
of the Dirac equation is needed.

Of course, the average over the gauge fields can be interchanged
with the one over the field $\varphi$ and one may replace 
(\ref{eq_gbarferm1}) by an average over a finite number of fields 
$\varphi$ drawn from a distribution satisfying (\ref{eq_gbarferm2}). 
We found that the fluctuations of such a noisy (unbiased) estimator for
${\rm d} S_{\rm f}^{\rm eff}/{\rm d} \eta$ are small compared to the ones of
${\rm d} S_{\rm g}/{\rm d} \eta$, already when only one field $\varphi$ 
(from a Gaussian distribution) is used per gauge configuration.  
This has hence been our method of choice in all simulations.

\section{Detailed numerical results \label{sect:res}}

The following tables list detailed parameters and results
of our simulations.

\clearpage

\begin{table}[htb]
  \centering\small
  \begin{tabular}[c]{lllccccr@{.}lc}\Hline\\[-2.0ex]
    $L/a$  & $\beta$ & $\kappa$ & $\gbar^2$ & $\Delta(\gbar^2)$ & $\vbar$ &
    $\Delta(\vbar)$ & \multicolumn{2}{c}{$m$} & $\Delta(m)$ \\[0.5ex]
    \Hline\\[-2.5ex]
\multicolumn{10}{l}{$u=0.9793$}\\ \hline\hline
 4 & 9.2364 & 0.1317486  & 0.9793  & 0.0007  & 0.1557  & 0.0019 & $-0$&$00600$ & 0.00011
\\
 5 & 9.3884 & 0.1315391  & 0.9794  & 0.0009  & 0.1322  & 0.0023 & $0$&$00197$ & 0.00005
\\
 6 & 9.5000 & 0.1315322  & 0.9793  & 0.0011  & 0.1266  & 0.0016 & $-0$&$00014$ & 0.00003
\\
 8 & 9.7341 & 0.131305   & 0.9807  & 0.0017  & 0.1177  & 0.0042 & $0$&$00074$ & 0.00006
\\ \hline
 8 & 9.2364 & 0.1317486  & 1.0643  & 0.0034  & 0.1244  & 0.0061 & $0$&$00010$ & 0.00004
\\       
10 & 9.3884 & 0.1315391  & 1.0721  & 0.0039  & 0.1151  & 0.0077 & $0$&$00210$ & 0.00003
\\       
12 & 9.5000 & 0.1315322  & 1.0802  & 0.0044  & 0.1227  & 0.0072 &$-0$&$00091$ & 0.00002
\\       
16 & 9.7341 & 0.131305   & 1.0753  & 0.0055  & 0.1047  & 0.0080 &$-0$&$00008$ & 0.00003
\\ \hline\hline
\multicolumn{10}{l}{$u=1.1814$}\\ \hline\hline
 4 & 8.2373 & 0.1327957  & 1.1814  & 0.0005  & 0.1483  & 0.0016 & $0$&$00100$ & 0.00011
\\             
 5 & 8.3900 & 0.1325800  & 1.1807  & 0.0012  & 0.1353  & 0.0018 & $-0$&$00018$ & 0.00009
\\             
 6 & 8.5000 & 0.1325094  & 1.1814  & 0.0015  & 0.1269  & 0.0014 & $-0$&$00036$ & 0.00003
\\             
 8 & 8.7223 & 0.1322907  & 1.1818  & 0.0029  & 0.1141  & 0.0048 & $-0$&$00115$ & 0.00004
\\ \hline      
 8 & 8.2373 & 0.1327957  & 1.3154  & 0.0055  & 0.1209  & 0.0061 & $0$&$00020$ & 0.00005
\\             
10 & 8.3900 & 0.1325800  & 1.3287  & 0.0059  & 0.1128  & 0.0070 & $0$&$00097$ & 0.00007
\\
12 & 8.5000 & 0.1325094  & 1.3253  & 0.0067  & 0.1304  & 0.0068 & $-0$&$00102$ & 0.00002
\\             
16 & 8.7223 & 0.1322907  & 1.3347  & 0.0061  & 0.1065  & 0.0049 & $-0$&$00194$ & 0.00002
\\ \hline\hline     
\multicolumn{10}{l}{$u=1.5031$}\\ \hline\hline       
 4 & 7.2103 & 0.1339411  & 1.5031  & 0.0010  & 0.1437  & 0.0010 & $-0$&$00074$ & 0.00010 
\\         
 5 & 7.3619 & 0.1339100  & 1.5044  & 0.0027  & 0.1250  & 0.0031 & $0$&$00052$ & 0.00010
\\         
 6 & 7.5000 & 0.1338150  & 1.5031  & 0.0025  & 0.1201  & 0.0024 & $-0$&$00078$ & 0.00004
\\ \hline  
 8 & 7.2103 & 0.1339411  & 1.7310  & 0.0059  & 0.1151  & 0.0037 & $0$&$00959$ & 0.00004
\\
10 & 7.3619 & 0.1339100  & 1.7581  & 0.0113  & 0.1062  & 0.0084 & $0$&$00257$ & 0.00005
\\         
12 & 7.5000 & 0.1338150  & 1.7449  & 0.0119  & 0.1223  & 0.0073 & $-0$&$00138$ & 0.00004
\\ \hline\hline     
\multicolumn{10}{l}{$u=1.7319$}\\ \hline\hline
 4 & 6.7251 & 0.1347424  & 1.7319  & 0.0020  & 0.1378  & 0.0009 & $-0$&$00181$ & 0.00013
\\             
 5 & 6.8770 & 0.1346900  & 1.7333  & 0.0032  & 0.1272  & 0.0025 & $-0$&$00005$ & 0.00011
\\             
 6 & 7.0000 & 0.1345794  & 1.7319  & 0.0034  & 0.1161  & 0.0023 & $-0$&$00002$ & 0.00005
\\ \hline      
 8 & 6.7251 & 0.1347424  & 2.0583  & 0.0070  & 0.1008  & 0.0032 & $0$&$01051$ & 0.00005
\\             
10 & 6.8770 & 0.1346900  & 2.0855  & 0.0208  & 0.0934  & 0.0080 & $0$&$00335$ & 0.00006
\\             
12 & 7.0000 & 0.1345794  & 2.0575  & 0.0196  & 0.0833  & 0.0082 & $-0$&$00049$ & 0.00006
\\ \Hline
\end{tabular}
  \caption{\fs Simulation parameters and results using the 1-loop value
    for $c_{\rm t}$.}
  \label{tab:data1}
\end{table}

\clearpage

\begin{table}[htb]
  \centering\small
\begin{tabular}[c]{lllccccr@{.}lc}\Hline\\[-2.0ex]
    $L/a$  & $\beta$ & $\kappa$ & $\gbar^2$ & $\Delta(\gbar^2)$ & $\vbar$ &
    $\Delta(\vbar)$ & \multicolumn{2}{c}{$m$} & $\Delta(m)$ \\[0.5ex]
    \Hline\\[-2.5ex]
\multicolumn{10}{l}{$u=1.5031$}\\ \hline\hline
 4 & 7.2811 & 0.1338383  & 1.5031  & 0.0012  & 0.1434  & 0.0013 & $0$&$00043$ & 0.00015
\\              
 5 & 7.4137 & 0.1338750  & 1.5033  & 0.0026  & 0.1310  & 0.0027 & $-0$&$00083$ & 0.00009
\\              
 6 & 7.5457 & 0.1337050  & 1.5031  & 0.0030  & 0.1247  & 0.0031 & $0$&$00072$ & 0.00008
\\              
 8 & 7.7270 & 0.133488   & 1.5031  & 0.0035  & 0.1219  & 0.0032 & $-0$&$00084$ & 0.00003
\\ \hline       
 8 & 7.2811 & 0.1338383  & 1.7204  & 0.0054  & 0.1091  & 0.0037 & $0$&$00928$ & 0.00004
\\              
10 & 7.4137 & 0.1338750  & 1.7372  & 0.0104  & 0.1110  & 0.0063 & $0$&$00109$ & 0.00004
\\              
12 & 7.5457 & 0.1337050  & 1.7305  & 0.0122  & 0.0893  & 0.0079 & $-0$&$00006$ & 0.00008
\\              
16 & 7.7270  & 0.133488   & 1.7231  & 0.0151  & 0.1017  & 0.0092 & $-0$&$00154$ & 0.00019
\\ \hline\hline 
\multicolumn{10}{l}{$u=2.0142$}\\ \hline\hline
 4 & 6.3650 & 0.1353200  & 2.0142  & 0.0024  & 0.1349  & 0.0017 & $0$&$00000$ & 0.00023
\\              
 5 & 6.5000 & 0.1353570  & 2.0142  & 0.0044  & 0.1236  & 0.0024 & $0$&$00002$ & 0.00011
\\              
 6 & 6.6085 & 0.1352600  & 2.0146  & 0.0056  & 0.1205  & 0.0029 & $0$&$00030$ & 0.00009
\\              
 8 & 6.8217 & 0.134891   & 2.0142   & 0.0102   & 0.0991  & 0.0045 & $0$&$00049$ & 0.00007
\\ \hline       
 8 & 6.3650 & 0.1353200  & 2.4814  & 0.0172  & 0.1016  & 0.0049 & $0$&$01318$ & 0.00008
\\              
10 & 6.5000 & 0.1353570  & 2.4383  & 0.0188  & 0.0900  & 0.0050 & $0$&$00367$ & 0.00005
\\              
12 & 6.6085 & 0.1352600  & 2.5077  & 0.0259  & 0.1074  & 0.0069 & $0$&$00013$ & 0.00004
\\              
16 & 6.8217 & 0.134891   & 2.475  & 0.031  & 0.0916  & 0.0076 & $-0$&$00053$ & 0.00006
\\ \hline\hline 
\multicolumn{10}{l}{$u=2.4792$}\\ \hline\hline
 4 & 5.8724 & 0.1360000  & 2.4792  & 0.0034  & 0.1206  & 0.0016 & $0$&$00000$ & 0.00026
\\              
 5 & 6.0000 & 0.1361820  & 2.4792  & 0.0073  & 0.1085  & 0.0023 & $-0$&$00009$ & 0.00014
\\              
 6 & 6.1355 & 0.1361050  & 2.4792  & 0.0082  & 0.1025  & 0.0032 & $0$&$00000$ & 0.00013
\\              
 8 & 6.3229 & 0.1357673  & 2.4792  & 0.0128  & 0.1015  & 0.0053 & $0$&$00000$ & 0.00016
\\ \hline       
 8 & 5.8724 & 0.1360000  & 3.2511  & 0.0277  & 0.0859  & 0.0043 & $0$&$01819$ & 0.00011
\\              
10 & 6.0000 & 0.1361820  & 3.3356  & 0.0502  & 0.0796  & 0.0064 & $0$&$00579$ & 0.00009
\\               
12 & 6.1355 & 0.1361050  & 3.1558  & 0.0552  & 0.0801  & 0.0079 & $0$&$00078$ & 0.00007
\\               
16 & 6.3229 & 0.1357673  & 3.3263  & 0.0472  & 0.0806  & 0.0074 & $0$&$00039$ & 0.00017
\\ \hline\hline  
\multicolumn{10}{l}{$u=3.3340$}\\ \hline\hline
 4 & 5.3574 & 0.1356400  & 3.3340  & 0.0109  & 0.1087  & 0.0013 & $0$&$00000$ & 0.00040
\\              
 5 & 5.5000 & 0.1364220  & 3.3340  & 0.0182  & 0.0965  & 0.0018 & $-0$&$00004$ & 0.00017
\\              
 6 & 5.6215 & 0.1366650  & 3.3263  & 0.0196  & 0.0894  & 0.0031 & $0$&$00051$ & 0.00018
\\              
 8 & 5.8097 & 0.1366077  & 3.334   & 0.019   & 0.0887  & 0.0042 & $0$&$00000$  & 0.00004
\\ \hline       
 8 & 5.3574 & 0.1356400  & 5.588   & 0.049   & 0.0576  & 0.0036 & $0$&$03163$ & 0.00019
\\
10 & 5.5000 & 0.1364220  & 5.430   & 0.098   & 0.0585  & 0.0048 & $0$&$01126$ & 0.00012
\\              
12 & 5.6215 & 0.1366650  & 5.624   & 0.089   & 0.0607  & 0.0043  & $0$&$00334$ & 0.00015
\\              
16 & 5.8097 & 0.1366077  & 5.4763  & 0.1236  & 0.0689  & 0.0052 & $0$&$00048$ & 0.00004
\\ \Hline 
  \end{tabular}
  \caption{\fs Simulation parameters and results using the 2-loop value
    for $c_{\rm t}$.}
  \label{tab:data2}
\end{table}
\vfill \eject

\clearpage

\end{appendix}

\bibliography{lattice}
\bibliographystyle{h-elsevier3}

\end{document}